\documentclass[hidelinks,onefignum,onetabnum]{siamart220329}
\usepackage[utf8]{inputenc}

\usepackage{cite}
\usepackage{amsmath,amssymb,amsfonts}

\usepackage{algorithm}
\usepackage[noend]{algpseudocode}

\usepackage{graphicx}
\usepackage{subcaption}
\graphicspath{figures}
\DeclareGraphicsExtensions{.pdf}
\usepackage{textcomp}
\usepackage{xcolor}

\usepackage{array}

\usepackage{url}

\usepackage{ltablex}
\usepackage{booktabs}
\usepackage{multirow}
\usepackage{etoolbox, siunitx}
\robustify\bfseries
\newrobustcmd\B{\DeclareFontSeriesDefault[rm]{bf}{b}\bfseries}
\usepackage{xspace}

\usepackage{bm}

\usepackage{enumitem}

\usepackage{comment}

\usepackage{hyperref}

\begin{document}
\title{Jet: Multilevel Graph Partitioning on \\Graphics Processing Units}

\author{Michael S. Gilbert\thanks{Pennsylvania State University, University Park, USA 
  (\email{msg5334@psu.edu}).}
\and Kamesh Madduri\thanks{Pennsylvania State University, University Park, USA 
  (\email{madduri@psu.edu}).}
\and Erik G. Boman\thanks{Sandia National Laboratories, Albuquerque, USA
    (\email{egboman@sandia.gov}).}
\and Sivasankaran Rajamanickam\thanks{Sandia National Laboratories, Albuquerque, USA
    (\email{srajama@sandia.gov}).}
}

\maketitle

\begin{abstract}
The multilevel heuristic is the dominant strategy for high-quality sequential and parallel graph partitioning. Partition refinement is a key step of multilevel graph partitioning. In this work, we present Jet, a new parallel algorithm for partition refinement specifically designed for Graphics Processing Units (GPUs). We combine Jet with GPU-aware coarsening to develop a $k$-way graph partitioner, the Jet partitioner. The new partitioner achieves superior quality compared to state-of-the-art shared memory partitioners on a large collection of test graphs.
\end{abstract}

\section{Introduction}

Parallel graph partitioning~\cite{BMS16} is a key enabler for both large-scale graph analytics~\cite{LNP16, Sakr21} and high-performance scientific computing~\cite{bisseling2020parallel, pothen1997graph}. Graph partitioning is the task of creating approximately equally sized disjoint sets of vertices in the graph, while simultaneously minimizing the \emph{cutsize}, the number of edges connecting vertices in different sets. Most graph partitioning software tools and algorithms use the \emph{multilevel} heuristic. The multilevel heuristic constructs a sequence of progressively smaller graphs in a \emph{coarsening} phase, finds a solution to the problem (partitioning in this case) on the smallest graph and then \emph{ not coarsening} the solution to fit the top-level graph. The uncoarsening step also \emph{improves} the solution using information from each graph in the sequence in a process called \emph{refinement}. Refinement algorithms for graph partitioning work by \emph{moving vertices} to improve the quality of the solution. The graph partition refinement problem is well studied in the context of shared-memory algorithms for multicore systems~\cite{mtMetis2016icpp, ASS20, kaminpar}. Our work considers the problem of partition refinement on graphics processing units (GPUs), with a focus on matching or exceeding partition quality obtained with fast multicore partitioners.


Our refinement algorithm, named Jet, decouples the two primary tasks of refinement algorithms: improving the cutsize and maintaining a balanced solution. This enables our algorithm to move larger sets of vertices without relying too much on fine-grained synchronization. This is critical for obtaining a high degree of parallelism. Moreover, we utilize a novel heuristic for selecting a set of vertices to move during the cutsize reduction phase. The heuristic enables our refinement to move larger sets of vertices in each pass, and to escape local minima. In this heuristic, we use information from the current partition state to assign a priority value to each vertex, and then approximate the expected value for the next partition state from these priorities. The expected value of the next partition state in the neighborhood of each vertex determines whether the vertex should move. On the majority of test graphs we experiment with, this heuristic results in higher-quality partitions than other parallel refinement schemes.

We develop a partitioner for GPUs utilizing our previous work on coarsening~\cite{GAB21} and Jet, our novel refinement algorithm. GPU acceleration enables our partitioner to achieve consistently faster partitioning times compared to other partitioners. Our partitioner also achieves consistently smaller cutsizes on graphs from varied domains such as finite element methods, social networks, and semiconductor simulations.

The following are the key algorithmic contributions and performance highlights:
\begin{itemize}[leftmargin=*]
\item We present Jet, a novel hiqh-quality, GPU-parallel, $k$-way refinement algorithm. Our experiments indicate that Jet outperforms the Multitry Local Search algorithm in terms of graph partition quality. 
\item We present a $k$-way graph partitioner, the Jet partitioner, that leverages GPU acceleration to attain 2$\times$ faster partitioning times than competing methods in a majority of test cases. We also modify the GPU implementation to adapt it for multicore execution.
\item We demonstrate superior quality when compared to state-of-the-art shared memory partitioners on a diverse test set of over 60 graphs.
\end{itemize}

\section{Background and Prior Work}
\label{s:priorwork}

\subsection{Problem Definition}
\label{ss:probdef}

Consider a graph $G$ with $n$ vertices (or nodes) and $m$ edges. We assume the graph is undirected and has no self-loops or parallel edges. Vertices can have associated positive integral weights. Edge weights are positive integers representing the strength of the connection of two vertices. Vertex-weight pairs are denoted by $V$, and weighted edge triples by $E$. For a positive integer $k$, a $k$-way partition of $G$ is a set of pairwise disjoint subsets of $V$ (or \emph{parts} $\{p_1, p_2, \ldots, p_k\} = P$) such that $\cup_{i = 1}^{k}p_i = V$. The weight/size of a part $p_i$ is the sum of the weights of its constituent vertices. Given a partition, the \emph{cut set} is the set of edges $\langle u, v, w_{uv}\rangle \in E$ with $u$ and $v$ in different parts. The sum of the weights of edges in the cut set is called the \emph{cost} (or \emph{cutsize}, or \emph{edge cut} in case of unweighted graphs) of the partition. A \emph{balance constraint} in the form of a non-negative real constant $\lambda$ places a limit on the part weights: $\mathrm{weight}(p_i) \leq (1 + \lambda) \frac{\mathrm{weight}(V)}{k}$, $\forall$ $1 \leq i \leq k$; there is no lower bound on the size of a part. The value of $\lambda$ is typically 0.01--0.1 (or a 1--10\% allowed imbalance). The objective of the $k$-way graph partitioning problem is to minimize the \emph{cost} of the partition of $G$ while satisfying the balance constraint. The output of the partitioning problem is typically an array of size $n=|V|$ mapping vertices to the parts.

\subsection{Multilevel partitioning}

The multilevel heuristic~\cite{Teng99} is extensively used in large-scale graph analysis. Its applications include graph partitioning~\cite{chacoPaper,BS94,Metis}, clustering~\cite{GraClus,GG05,AB12}, drawing~\cite{ACE,HS15}, and representation learning~\cite{HARP,GOSH}. The family of algebraic multigrid methods~\cite{Brandt86,XZ17} and multilevel domain decomposition methods~\cite{Smith04,FROSch} in linear algebra are closely related to multilevel methods for graph analysis. In a multilevel method, instead of solving a problem on a large graph, we build a hierarchy of graphs that are progressively smaller than the original graph and yet preserve the structure of the original graph. We then solve the problem on the smallest graph and \emph{project} or \emph{interpolate} the solution to the original graph using the hierarchy. Algorithm~\ref{alg:MLpart} gives the high-level template for multilevel graph partitioning. 
\textsc{MLCoarsen} returns a sequence of coarser graphs and the corresponding vertex mappings.
\textsc{ProjectPartition} is a lightweight routine that copies the solution (partition array) from the previous level to the vertices at the current level. Refinement is applied after projection on each level to reduce the cutsize, and to satisfy the balance constraint if the coarser partition could not satisfy it.

\begin{algorithm}[htbp]
    \caption{A template for multilevel graph partitioning.}
    \label{alg:MLpart}
    \begin{algorithmic}[1]
    \Require graph $G$ as defined in Section~\ref{ss:probdef}, number of parts $k$, balance $\lambda$.
    \Ensure A partition array $P_0[0..n-1]$, where $P_0[v]$ indicates the partition that vertex $v \in V$ belongs to.
    \State $\{G_0, \ldots,G_l\}, \{M_0, \ldots,M_l\} \gets$ \textsc{MLCoarsen}($G$) 
    \State $P_l \gets$ \textsc{InitialPartition}($G_l, k, \lambda$)
    \State $P_l \gets $\textsc{RefinePartition}($G_l, P_l, k, \lambda$)
    \State $i \gets l-1$
    \While {$i \geq 0$} \Comment{$l$ \emph{Uncoarsening} steps}
        \State $P_i \gets$ \textsc{ProjectPartition}($P_{i+1}, M_{i+1}$)
        \State $P_i \gets$ \textsc{RefinePartition}($G_i, P_i, k, \lambda$)
        \State $i \gets i - 1$
    \EndWhile
    \end{algorithmic}
    \end{algorithm}    

Since there is a clear separation of multilevel coarsening, initial partitioning, and refinement in multilevel level partitioning, we focus on multilevel refinement in this work. Sequential and parallel algorithms for multilevel coarsening have been extensively studied~\cite{Metis,SSS15,Mongoose,GAB21}.

\subsection{GPU: Related Work}
Most graph partitioners are designed for CPUs and do not run on GPUs. In particular, the refinement step in the multilevel algorithm is difficult to parallelize on a GPU. 
The first GPU partitioner we know of was developed by Fagginger Auer and Bisseling \cite{FaggingerAuer2013}. They developed two algorithms for GPU: one multilevel spectral, and the other was multilevel with greedy refinement. Their code was never released. A later GPU partitioner \cite{GKSG-GPU} implemented a multilevel algorithm with a label propagation-based refinement algorithm.
Sphynx \cite{SPHYNX,Sphynx21} is a spectral partitioner that runs on GPU. It is not multilevel. Although it runs quite fast on GPU, the cut quality is significantly worse (up to $50\times$) than Metis/ParMetis on irregular graphs. Therefore, in this paper, we do not consider Sphynx any further.

\subsection{Refinement}

The objective of the partition refinement problem is identical to graph partitioning. Refinement algorithms improve an input partition; in the multilevel method, this partition is an output from coarser levels in the multilevel hierarchy. It can also be used outside the context of multilevel partitioning, regardless of the method used to produce the given partition. Refinement is \emph{local} in nature; information about the current partition state is used to generate the next state. Refinement methods frequently use a vertex attribute called the \emph{gain}, which is defined according to the current partition state. For bipartitioning, the gain describes the decrease in cutsize for moving a vertex from its current part to the other part. This quantity is negative if the cutsize would increase. When $k > 2$, we use \emph{gain} to indicate the expected decrease in cutsize for a \emph{vertex move} (moving a single vertex from its current part to a specific destination part).

Refinement algorithms typically operate in \emph{iterations} over a set of vertices. The set can either be all vertices, or a subset of vertices such as the \emph{boundary} set, or some other subset of interest. The number of iterations is typically a small constant, and it is desirable that the running time of one refinement pass be linear in $m=|E|$. A vertex $v$ is in the boundary set if there exists a vertex $u$ in the neighborhood of $v$ such that $\mathrm{part}(v) \neq \mathrm{part}(u)$. As no vertex outside the boundary set can have a positive gain vertex move, it is common for refinement iterations to exclusively consider this set.

\subsection{Refinement: Related Work}

In this work, we are interested in parallel partition refinement schemes. Several recent papers~\cite{mtMetis, mtMetisOpt, mtMetis2016icpp, ASS20, mtkahypar_kwayfm} have demonstrated that parallel refinement techniques can obtain a similar or better quality to the sequential refinement algorithms from which they are derived. We group algorithms into four broad categories and describe them below.

\subsubsection{Label Propagation}

Several refinement algorithms share similarities with an iteration of the Label Propagation (LP) community detection algorithm~\cite{raghavan2007near}. Thus, we group them into a common category. In these algorithms, the neighborhood of each vertex is examined to determine the part to which the vertex is most connected. The vertex is then moved to this part if doing so does not violate the balance constraint. A typical serial implementation visits each vertex of a graph at most once per iteration, in arbitrary order. Common orderings include random shuffles and the natural order of the vertices. More complex orderings make use of priority queues to determine the vertex which results in the largest decrease in cutsize. This technique cannot escape local minima, which occur when no single vertex can be moved for a decrease in cutsize without violating the balance constraint. In parallel implementations, each processor owns a subset of the vertices, and each processor visits the vertices it owns in some order. A parallel implementation is synchronous if there is a barrier synchronization after all vertices are inspected in an iteration, or termed asynchronous if part changes are immediately applied. 
The balance constraint can be maintained in a parallel setting by atomically updating the part sizes. 
Mt-Metis \cite{mtMetis}, mt-KaHIP \cite{ASS20}, KaMinPar \cite{kaminpar}, and Mt-KaHyPar \cite{mtkahypar_kwayfm} all implement variations in a multilevel setting as a refinement option or the primary refinement method. PuLP \cite{slota2014pulp} implements this technique for direct partitioning outside of a multilevel framework, using random initial partitions. The latest GPU partition refinement algorithm \cite{GKSG-GPU} that we know uses a synchronous scheme to fill a move buffer. It considers the top $x$ moves in the buffer at a time, and determines the best of all $2^x$ permutations of performing or not performing each move. This imposes a practical limit on the rate at which the move buffer can be processed.

\subsubsection{Localized FM Search}
Mt-KaHIP's multitry local search (MLS) \cite{ASS20} and Mt-KaHyPar's parallel k-way FM (KFM) \cite{mtkahypar_kwayfm, mtkahypar_kwayfm2} search for sequences of vertex moves that may begin with a negative gain move, but collectively improve the cutsize. These algorithms relax the well-known FM algorithm \cite{FM82}. Each algorithm begins multiple FM-style searches seeded from a small number of boundary vertices, whereas the standard FM performs a single search seeded from all boundary vertices. A local search repeatedly selects a vertex to move from the top of a priority queue, which is keyed by the maximum gain of moving each vertex to another part. Each vertex move requires inserting its neighborhood into the queue, or updating its neighbors already within the queue. A search ends when it has exhausted its queue or when a stopping condition is triggered. This stopping condition is based on the statistical likelihood that a search will yield further improvement. A single iteration of MLS or KFM begins from an unordered list of boundary vertices, and seeds many searches from this list until the list is empty. Multiple local searches can occur in parallel, up to the limit of available threads. At the end of an iteration, the vertex moves performed by each search are joined into a single sequence, and the best prefix of this sequence that satisfies the balance constraint is committed. MLS and KFM differ in terms of how many vertices are used to initialize each search, the visibility of vertex moves between concurrent searches, and how each search is concatenated into the global sequeuence. The MLS refinement in mt-KaHIP produces higher quality partitions than the fast and eco configurations of the serial partitioner KaHIP \cite{SS11} and the parallel partitioners Mt-Metis and ParHIP, according to the experiments of the authors \cite{ASS20}. Mt-KaHyPar-D (default configuration) using KFM refinement produces higher quality partitions than mt-KaHIP \cite{mtkahypar_kwayfm2}.

\subsubsection{Hill-Scanning}
Hill-scanning refinement \cite{mtMetis2016icpp} is another variant of localized FM search, except that each search immediately ends when achieving a net positive gain. The sequence of moves built by a search is termed as a \emph{hill}, and hills can't grow beyond a maximum size (16 vertices within Mt-Metis). Hills which attain positive total gain are applied to the partition, otherwise they are discarded. The most significant difference between hill-scanning and MLS/KFM is the elimination of any need to revert moves. Hill-scanning exploits parallelism by statically dividing the vertices among the processors, but a processor that is building a hill can use vertices owned by another processor. In this way, hills may overlap.
Overlap between two hills can be corrected in successive iterations, but this may not happen if doing so would violate the balance constraint. A serial implementation of the hill-scanning technique was shown to attain similar or superior quality to other serial refinement schemes \cite{mtMetis2016icpp} including Fiduccia-Mattheyses (FM) with recursive bisection, k-way pairwise FM, and Multi-try FM (a weaker precursor to MLS and KFM). The cutsizes produced by hill-scanning degrade by about 0.5\% when run with 24 threads instead of serially \cite{mtMetis2016icpp}. The authors of mt-KaHIP \cite{ASS20} found that hill-scanning as implemented in Mt-Metis has substantial difficulty maintaining the balance constraint when the number of processors is large.

\subsubsection{Network Flow Methods}
Max-flow min-cut solvers have seen great success as partition refinement algorithms \cite{SS11, parFlow, mtkahypar_kwayfm2}. Mt-KaHyPar-Q (high quality configuration) creates a network flow problem by growing a region around the boundary between two parts. It uses a parallel implementation of the push-relabel algorithm to compute a minimum cut inside this region, and this new cut replaces the old cut if it satisfies the balance constraint. While flow-based methods outperform other refinement methods in terms of result quality, they are also considerably more expensive. 

\section{Our Partitioner}

We now discuss our new multilevel GPU partitioner with an emphasis on the partition refinement algorithm. We coarsen until the coarsest graph obtained is extremely small, typically between $4k$ and $8k$ vertices. We use the $k$-way partitioning method in Metis \cite{Metis} to perform the initial partitioning. Since the coarsest graph is very small, GPU parallelization of the initial partitioning is left for future work.


\subsection{Coarsening}

Our coarsening approach is based on a GPU implementation discussed in \cite{GAB21}, specifically the \emph{two-hop matching} approach originally developed for the Mt-Metis partitioner~\cite{mtMetisOpt}. This approach begins with a standard heavy-edge matching and only adds two-hop matches if more than 25\% of all vertices are unmatched.

Two-hop matchings can be split into three categories: leaves, twins, and relatives. A pair of vertices are relatives if they are separated by a distance of two in the graph. Twins are a subset of relatives where the neighborhoods of both vertices are the same. Leaves are a subset of twins that have degree one. We extended our previous work in two ways. First, we use a hashing scheme to perform twin matching. Second, we implement relative matching using \emph{matchmaker vertices}. Matchmaker vertices are matched vertices with unmatched neighbors, and matches are performed within the neighborhoods of these matchmakers. We exclude vertices with very high degree from acting as matchmakers. We have also replaced our contraction scheme from our previous work with a fine-grained per-vertex hashing scheme for deduplication, as outlined in Algorithm~\ref{alg:contraction}.

\begin{algorithm}[htbp]
    \caption{Edge Contraction Algorithm.}
    \label{alg:contraction}
    \begin{algorithmic}[1]
    \Require The graph $G = (V, E)$ as defined in Section~\ref{ss:probdef}. The mapping vector $C$. Coarse vertex count $n_c$.
    \Ensure The coarse edges $E_c$.
    \State bound $\gets$ zeros($n_c$)
    \For{$v \in V$ in parallel}
        \State bound[$C[v]$] $\gets$ bound[$C[v]$] + $|E[v]|$
    \EndFor
    \State offsets $\gets $ exclusivePrefixSum(bound)
    \State $H_{key} \gets $ nulls($|E|$) \Comment{Initialize per-vertex hash table}
    \State $H_{val} \gets $ zeros($|E|$)
    \For{$v \in V$ in parallel}
        \State $v_c \gets C[v]$
        \State $h_{key} \gets $ $H_{key}$[offsets[$v_c$]..offsets[$v_c+1$]] \Comment{Hash table to use}
        \State $h_{val} \gets $ $H_{val}$[offsets[$v_c$]..offsets[$v_c+1$]]
        \For{$(u,w) \in E[v]$ in parallel}
            \State $u_c \gets C[u]$
            \State $i \gets $ insertOrLookup($h_{key}$, $u_c$)
            \State $h_{val}[i]$ $\gets$ $h_{val}[i]$ + $w$
        \EndFor
    \EndFor
    \State $E_c \gets $ extractInsertions(H, Hv)
    \end{algorithmic}
    \end{algorithm}

\subsection{Kokkos}
We use Kokkos~\cite{Kokkos} to implement the parallel kernels in our code. Kokkos facilitates performance portability, allowing the programmer to maintain a single-source program that can be compiled for different shared-memory architectures. We compile for three different targets: NVIDIA GPUs using the CUDA backend, multicore CPUs using the OpenMP backend, and single threads of the same CPUs using a serial backend. The Kokkos programming model involves expressing a task as a sequence of small kernels that fit one of three parallel primitives: parallel-for, reduction, and scan.

\section{Jet Refinement Algorithm}

We have two design goals for refinement on the GPU: matching or exceeding the quality of multicore refinement techniques, and running time that is comparable to fast multicore refinement. Prior shared-memory multicore-centric refinement algorithms such as hill-scanning and MLS rely on thread-local priority queues. Priority queues are useful for finding sequeunces of moves that improve the cutsize where single moves cannot. However, these priority queue operations do not expose adequate concurrency for GPU-scale parallelism, and  therefore such approaches are not viable on the GPU. Size-constrained LP-based iterations can visit the vertices in any order, and therefore lends itself naturally to both multicore and GPU parallelism. However, the size constraint limits the number of vertices that can be moved in each pass. This can be especially problematic if the distribution of beneficial moves is biased towards certain destination parts. To address this challenge, our method, Jet, splits a size-constrained LP iteration into two phases. The first phase is an unconstrained LP phase, Jetlp, that performs vertex moves while ignoring size constraints. The second phase is a rebalancing phase, Jetr, which has the task of moving vertices from oversized parts to non-oversized parts such that no oversized parts remain. It is paramount for the rebalancing phase to minimize any increase in cutsize (or loss). LP-based algorithms generally produce lower-quality results than FM-based methods, so we introduce novel augmentations to LP for improved quality. The overall structure of our refinement algorithm (Algorithm~\ref{alg:refinement}) is to apply Jetlp until any part becomes oversized, then apply Jetr until balance is restored. We denote each application of either Jetlp or Jetr as an ``iteration''. We record the best balanced partition in terms of cutsize, and terminate refinement when we exceed a certain number of iterations (we use 12 for our results) without encountering a new best partition. We also use a tolerance factor $\phi$ to terminate when the cutsize is improving too slowly (see line \ref{alg:tol_check}). $\phi$ is the most important hyper-parameter to control the quality/runtime tradeoff, where $\phi = 1$ gives the best quality. We use $\phi = 0.999$, which gives a good balance between quality and runtime. 

\begin{algorithm}[htbp]
\caption{Jet Refinement Algorithm.}
\label{alg:refinement}
\begin{algorithmic}[1]
\Require The graph $G = (V, E)$ as defined in Section~\ref{ss:probdef}. The number of parts $k$, balance factor $\lambda$. A partition array $P_0$.
\Ensure An output partition array $P_{best}$.
\State $P_{best} \gets P_0$
\State $P_{iter} \gets P_0$
\State $DS \gets $ initDataStructures($G, P_0, k$)
\State $R \gets \emptyset$
\While {iteration limit not reached}
    \If{imb($G,P_{iter},k) <\lambda$}
        \State $M \gets $ Jetlp($G,P_{iter},DS,R$)
        \State $R \gets $ vertexSet($ML$)
        \State reset weak rebalance counter
    \Else
        \If{weak rebalance limit not reached}
            \State $M \gets$ Jetrw($G,P_{iter},DS,k,\lambda$)
        \Else
            \State $M \gets$ Jetrs($G,P_{iter},DS,k,\lambda$)
        \EndIf
    \EndIf
    \State $P_{iter}, DS \gets $ updatePartsAndDS($G,P_{iter},k,M, DS$)
    \If{imb($G,P_{iter},k) <\lambda$}
        \If{cost($G,P_{iter}) <$ cost($G,P_{best}$)}
            \If{cost($G,P_{iter}) <\phi*$cost($G,P_{best}$)}
            \label{alg:tol_check}
                \State reset iteration counter
            \EndIf
            \State $P_{best} \gets P_{iter}$
        \EndIf
    \ElsIf{imb($G,P_{iter},k) <$ imb($G,P_{best},k$)}
        \State $P_{best} \gets P_{iter}$
        \State reset iteration counter
    \EndIf
\EndWhile
\end{algorithmic}
\end{algorithm}

\subsection{Unconstrained Label Propagation - Jetlp}

Our unconstrained label propagation is synchronous, i.e., updates to the partition state are deferred to the end of each iteration. The steps in Algorithm~\ref{alg:jetLP} are as follows: first, the algorithm selects a destination part $P_d(v)$ for each vertex $v$, and records the gain $F(v)$ of making this move by itself. Second, it filters the vertices where $P_d(v)$ is different from the current part $P_s(v)$ and the gain $F(v)$ satisfies a constraint (equation \ref{eq:gainconn_filter}); it pushes these vertices to an unordered list, and assigns the gain as the priority value. Finally, it filters this unordered list using an approximation of the expected value of the next partition state. It determines this approximation in the neighborhood of each vertex that passed the first filter, by merging $P_s$ and $P_d$ according to the priority values within each neighborhood. It commits all moves that pass the second filter, and then updates the data structures that track connectivity of each vertex and the sizes of each part. The name Jet derives from a similarity in structure to a jet engine: the selection of destination parts is similar to the compressor, the first filter to the combustion chamber, and the second filter to the afterburner.

\begin{algorithm}[htbp]
\caption{Jet - Label Propagation (Jetlp).}
\label{alg:jetLP}
\begin{algorithmic}[1]
\Require The graph $G = (V, E)$. Partition array $P_s$. Data structures $DS$ for querying vertex-part connection info. Locked vertices $R \subset V$. Filter ratio $c$.
\Ensure A list of moves $M$, in the form of vertex-destination part pairs.
\State $P_d \gets P_s$
\State $F \gets$ negativeInfinity($|V|$)
\For{$v \in V \setminus R$ in parallel}
    \State $A_v \gets \textnormal{adjacentParts}(v, DS) \setminus \{P_s[v]\}$
    \If{$A_v \neq \emptyset$}
        \State $P_d[v] \gets $ argmax$_{p \in A_v}$conn($v, p, DS$)
        \State $F[v] \gets $ conn($v, P_d[v], DS$) $-$ conn($v, P_s[v], DS$)
    \EndIf
\EndFor
\State $X \gets $ gainConnRatioFilter($V \setminus R, P_s, F, DS, c$) \Comment{First filter (Equation \ref{eq:gainconn_filter})}
\State $F_2 \gets $ zeros($|X|$)
\For{$v \in X$ in parallel}
    \For{$(u,w) \in E[v]$ in parallel}
        \State $p_u \gets P_s[u]$
        \If{$\mathit{ord}(u) < \mathit{ord}(v)$}
            \State $p_u \gets P_d[u]$
        \EndIf
        \If{$p_u = P_d[v]$}
            \State $F_2[v] \gets F_2[v] + w$
        \ElsIf{$p_u = P_s[v]$}
            \State $F_2[v] \gets F_2[v] - w$
        \EndIf
    \EndFor
\EndFor
\State $M \gets $ nonNegativeGainFilter($X, P_d, F_2$) \Comment{Second filter}
\end{algorithmic}
\end{algorithm}

\subsubsection{Changes to address LP limitations}

A synchronous implementation of LP-based refinement has two limitations. First, it is not possible to improve cutsize through negative gain vertex moves. Second, vertex moves in the same iteration can affect each other detrimentally. We introduce a method to address both of these problems: the \emph{vertex afterburner}. The vertex afterburner is a heuristic-based conflict resolution scheme permitting negative-gain vertex moves. We use the term afterburner as it is a secondary filter on the list of possible vertex moves; in typical LP-based refinement algorithms, there is only the first filter. Given a list of potential vertex moves $X$, we recompute the gain for each vertex in $X$ according to an approximation of the next partition state in its neighborhood. This approximation is created by merging $P_s$ with $P_d$, using an ordering $\mathit{ord}$. Due to the ordering $\mathit{ord}$, the approximations generated for overlapping neighborhoods are not consistent. $P_d$ is fixed for all vertices in $X$ prior to applying the afterburner, therefore recomputing the gain for each vertex $v \in X$ only involves the parts $P_d(v)$ and $P_s(v)$ specific to the move. For each neighbor $u$ of a vertex $v \in X$, if $\mathit{ord}(u) < \mathit{ord}(v)$, we calculate $v$'s gain assuming $u$ will move to $P_d(u)$. Otherwise, we assume $u$ remains in $P_s(u)$. This allows for vertex moves which initially had negative gain to become positive gain, and vice versa, depending on the other moves in $X$. The final move list $M$ is chosen as a subset of $X$, containing only the moves in $X$ with non-negative gain after recalculation. Let $F(x) = \textnormal{conn}(x, P_d(x)) - \textnormal{conn}(x, P_s(x))$ be the priority values for each vertex move, given by the gain values of each vertex move in a vacuum. $\mathit{ord}$ is defined as follows:
\begin{equation}
\begin{cases}
\mathit{ord}(u) < \mathit{ord}(v) & u \in X \land F(u) > F(v) \\
\mathit{ord}(u) < \mathit{ord}(v) & u \in X \land F(u) = F(v) \land u < v \\
\mathit{ord}(u) > \mathit{ord}(v) & \textnormal{otherwise}
\end{cases}
\end{equation}

\subsubsection{Negative Gain Moves}
\label{sec:neg_gain}
The efficacy of this filter heuristic is sensitive to the composition of $X$. If $X$ is selected too conservatively (i.e., only positive gain vertex moves), then afterburning does not produce an additional benefit over standard LP. If $X$ is not constrained (i.e. the entire boundary vertex set), then afterburning will produce worse results than standard LP. To determine the composition of $X$, we must first determine $P_d(v)$ for each vertex $v$:
\begin{equation}
    P_d(v) = \textnormal{argmax}_{p \in P \setminus \{P_s(v)\}} \textnormal{conn}(v, p)
\end{equation}
If a vertex is only connected to $P_s(v)$, it is not a boundary vertex and therefore is always excluded from $X$. The primary criterion for a vertex to be selected into $X$ is as follows:
\begin{equation}
\label{eq:gainconn_filter}
    \texttt{-}F(v) < \lfloor(c)\textnormal{conn}(v, p_s)\rfloor \lor F(v) \geq 0
\end{equation}
$c$ is a constant that can be adjusted for different levels of the multilevel hierarchy. We find experimentally that $c = 0.25$ is most effective for the finest level of the hierarchy, whereas $c = 0.75$ is best for all other levels (for our partitioner). It is important to note the floor rounding, as our results on certain graphs are sensitive to the rounding direction. We find that the coarsening and initial partitioning algorithms affect the optimal choice for $c$. 

\subsubsection{Vertex Locking}
We employ an additional technique that is intended to help migrate the boundary in a coordinated fashion over successive iterations. This technique uses a lock bit, which excludes all vertices selected in $M$ by an iteration of Jetlp from being chosen into $X$ in the next iteration of Jetlp. Locking helps to prevent oscillations, which occur when a vertex moves back and forth between two parts in successive Jetlp iterations. These oscillations may decrease solution quality by increasing the difficulty in changing the boundary's shape and location. Locks do not affect rebalancing iterations, nor does rebalancing change the lock state of any vertex.

\subsection{Rebalancing - Jetr}
\subsubsection{Two parts}

We introduce rebalancing with a simpler version applicable only when $k = 2$. Without loss of generality, let $p_a$ be the overweight part, and let $p_b$ be the other part. The goal of our rebalancing is to move the vertices from $p_a$ to $p_b$ until $p_a$ is no longer overweight, while minimizing the increase in the cutsize. We assign a simple loss value to every vertex in $p_a$: $\textnormal{loss}(v) = \textnormal{conn}(v, p_a) - \textnormal{conn}(v, p_b)$. Loss can also refer to the combined loss of vertex sets: $\textnormal{loss}(Z) = \sum_{v \in Z} \textnormal{loss}(v)$. We order the vertices of $p_a$ in terms of increasing loss in a list $L$. We then select the prefix $L_x$ of $L$ that minimizes the following equation:
\begin{equation}
\label{eq:prefix}
    |(|L_x| - (|p_a| - (1 + \lambda)|V|/k))|
\end{equation}
It is expensive to use a sort to obtain $L$, so we approximate $L$ with $L'$, which is sorted according to a partial ordering. This partial ordering is derived from the following function of the loss value:
\begin{equation}
\textnormal{slot}(x) = 
\begin{cases}
2 + \lfloor \log_2(x) \rfloor & x > 0 \\
1 & x = 0 \\
0 & x < 0
\end{cases}
\end{equation}
We found experimentally that the frequency of loss values tends to decrease as the absolute value of the loss value increases. We use $\log_2$ to assign slot values so that there are more slots closer to zero than far away from zero. This partial ordering is similar to a bucketing approach used to calculate the top k elements in a vector \cite{topKBucket}, but it only approximates the top-k elements to save time. The insertion order within each bucket is subject to race conditions. To reduce the atomic contention on GPU for the size counters of each bucket, we create $\rho$ sub-buckets within each bucket that are keyed by $v \mod \rho$. This bucket-oriented approach also integrates well when computing lists for multiple overweight parts independently in our $k > 2$ variations. \\ \\
\begin{theorem}
\label{th:loss}
Let $L'_x$ be the prefix of $L'$ that minimizes equation \ref{eq:prefix}. In a graph with uniform vertex weights, and assuming the number of vertices with negative loss is negligible, we have the following inequality:
\begin{equation}
\label{eq:loss}
    \textnormal{loss}(L'_x) \leq 2 \, \textnormal{loss}(L_x)
\end{equation}
\end{theorem}
We now prove this theorem. $|L'_x| = |L_x|$ because all vertices have the same weight. Theorem \ref{th:loss} holds trivially if both $L'_x$ and $L_x$ are the empty set. Otherwise, let $s=\max_{v \in L'_x}\textnormal{slot}(\textnormal{loss}(v))$. Let $S$ be the subset of $p_a$ consisting of all vertices with a slot value less than or equal to $s$. $L'_x$ is a subset of $S$ by definition. $S$ contains all the vertices in $p_a$ with loss values smaller than a function of $s$, therefore it is a prefix of $L$. $L_x$ must then be a subset of $S$, as $|S| \geq |L'_x| = |L_x|$. Similar logic shows that $S'$, the subset of all vertices in $p_a$ with slot values less than $s$, is a strict subset of both $L_x$ and $L'_x$. We have shown that $L_x \triangle L'_x$ is a subset of $S \setminus S'$. $L_x \triangle L'_x$ only contains vertices with loss values equal to $s$, by the definition of $S \setminus S'$. $|L'_x \setminus L_x| = |L_x \setminus L'_x|$ because $|L'_x| = |L_x|$. Any two vertices with the same slot value have loss values within a multiple of 2 of each other, therefore the following inequality is true:
\begin{equation}
    \textnormal{loss}(L'_x \setminus L_x) \leq 2 \, \textnormal{loss}(L_x \setminus L'_x)
\end{equation}
$\textnormal{loss}(L_x \cap L'_x) \geq 0$, $\textnormal{loss}(L'_x \setminus L_x) \geq 0$, and $\textnormal{loss}(L_x \setminus L'_x) \geq 0$ due to our assumption that there are negligible negative loss vertices. We can add $\textnormal{loss}(L_x \cap L'_x)$ to both sides to obtain equation \ref{eq:loss}. This inequality also holds for our k-way formulations. The assumption for uniform vertex weights is necessary to ensure $|L_x| = |L'_x|$. If we have non-uniform vertex weights, equation \ref{eq:loss} no longer holds. In this case, the ratio between the total number of vertices in each set having slot value $s$ (i.e., $|L_x \setminus S'|$ and $|L'_x \setminus S'|$) can be used to form a new inequality:
\begin{equation}
    \textnormal{loss}(L'_x) \leq 2 \, \frac{|L'_x \setminus S'|}{|L_x \setminus S'|} \, \textnormal{loss}(L_x)
\end{equation}

\subsubsection{More than two parts}

When $k > 2$, extending this rebalancing formulation is not trivial. We propose two separate extensions for arbitrary $k$ that both reduce to the $k = 2$ formulation. Similar to label propagation, the output consists of an unordered list of vertices to move and their chosen destinations. Let $B$ be the set of parts with size less than a value $\sigma$. $\sigma$ determines the maximum size for a part to be considered a valid destination and is chosen such that there is a deadzone between the size of valid destination parts and the size of oversized parts. The first formulation uses the following definition of loss:
\begin{equation}
\textnormal{loss}(v) = \max_{p_b \in B}{\textnormal{conn}(v, p_b)} - \textnormal{conn}(v, p_a)
\end{equation}
In this formulation (detailed in Algorithm \ref{alg:wrebalance}), vertices are evicted from the oversized parts such that each oversized part is just smaller than the size limit (this should be within the deadzone). This process is similar to the formulation with $k = 2$, except that there are multiple oversized parts. Note that the multiple scans performed from line \ref{alg:beginScan} to line \ref{alg:endScan} can be accomplished with just two scans, although we omit this detail from Algorithm \ref{alg:wrebalance} for brevity. The evicted vertices are sent to their best connected part among the valid destination parts. It is possible that the vertex is not connected to any valid destination part, in which case a random valid destination is chosen. In this formulation, it is possible for destination parts to become oversized. However, the deadzone prevents oversized parts from becoming valid destinations. This guarantees at most $k$ iterations to achieve a balanced partition as at least one part will move into the deadzone in each iteration, if the vertex weights are uniform. We observe that the typical number of iterations required is substantially less than $k$. We denote this extension as weak rebalancing (Jetrw) due to the potential need for many iterations.
\\ \\
Let $A_v$ be the adjacent parts of vertex $v$. Our second extension uses the following definition of loss:
\begin{equation}
\textnormal{loss}(v) = \textnormal{mean}_{p_b \in B \cap A_v}\textnormal{conn}(v, p_b) - \textnormal{conn}(v, p_a)
\end{equation}
Vertices are evicted from oversized parts in the same manner as the prior formulation. The destination parts then try to acquire as close to $\sigma - |B|$ vertices from the evicted set as possible. Given that the evicted vertices are arranged in an unordered list, each destination partition selects a contiguous group from this list. Destination partitions are overlayed onto the unordered list according to their capacity, forming a one-dimensional ``cookie-cutter'' pattern. This formulation guarantees that no oversized parts remain after a single iteration, if vertex weights are unit. We observe that vertex weights are often a significant fraction of the size constraint when more than one iteration is necessary. We denote this extension as strong rebalancing (Jetrs) due to its ability to achieve balance in one iteration in most scenarios.
\\ \\
Jetrw is much more effective at minimizing loss than Jetrs, even though it may require more iterations to converge upon a balanced partition. Our observations indicate that Jetrs requires fewer iterations to converge in any of the following scenarios: regular graphs, small values of $k$, and large imbalance ratios. We propose a combination of the two formulations, where we apply Jetrw for a certain number of iterations (denoted as $b_{max}$ in Algorithm~\ref{alg:refinement}), and then apply Jetrs if the partition is still unbalanced. We find that even a single iteration of Jetrw followed by an iteration of Jetrs can achieve much of the benefit of an unlimited number of iterations of Jetrw. Our full rebalancing (Jetr) consists of two iterations of Jetrw followed by a single iteration of Jetrs. If more iterations are necessary due to large vertex weights, these are performed with Jetrs. For both rebalancing variants, we find it beneficial to restrict a vertex from leaving an oversized partition if its respective vertex weight is greater than $1.5(|p_a| - \frac{|V|}{k})$. This restriction is applied before we construct $L'$.

\begin{algorithm}[htbp]
\caption{Jet - Weak Rebalancing.}
\label{alg:wrebalance}
\begin{algorithmic}[1]
\Require The graph $G = (V, E)$. An unbalanced partition array $P_s$. Data structures $DS$ for querying vertex-part connection info. $k$. $\lambda$. Minibucket count $\rho$.
\Ensure A list of moves $M$, in the form of pairs of vertex-destination parts.
\State $P_d \gets P_s$
\State $o \gets (1 + \lambda)|V|/k$ 
\State $\sigma \gets $ maxDestSize($o$)
\State $A \gets \{p \mid p \in P \land |p| > o\}$
\State $B \gets \{p \mid p \in P \land |p| \leq \sigma\}$
\State $F \gets$ zeros($|P_d|$)
\For{$v \in V$ in parallel}
    \If{$P_s[v] \in A$ and vtxWgt($v$) $<$ limit($P_s[v],|V|,k$)}
        \State $A_v \gets \textnormal{adjacentParts}(v, DS) \cap B$
        \State $P_d[v] \gets $ argmax$_{p \in A_v}$conn($v, p$)
        \If{$A_v = \emptyset$}
            \State $P_d[v] \gets $ randomPart($B$)
        \EndIf
        \State $F[v] \gets $ conn($v, P_s[v]$) $-$ conn($v, P_d[v]$)
    \EndIf
\EndFor
\State $L' \gets $ buckets($|A|$)
\For{$v \in V$ in parallel}
    \If{$P_s[v] \in A$}
        \State $s \gets \textnormal{slot}(F[v])$
        \State writeToBucket($L'[P_s[v]][s][v \textnormal{ mod } \rho], v$)
    \EndIf
\EndFor
\State $t \gets 0$
\State $M \gets $ emptyList
\For{$p_s \in A$ in parallel} \label{alg:beginScan}
    \State $m \gets 0$
    \State $m_{max} \gets |p_s| - o$
    \For{$v \in L'[p_s]$ parallel scan on $m$}
        \State $m \gets m + $vtxWgt($v$)
        \If{$m < m_{max}$}
            \State $M[t] \gets (v, P_d[v]$)
            \State $t \gets t + 1$
        \EndIf
    \EndFor
\EndFor \label{alg:endScan}
\end{algorithmic}
\end{algorithm}

\subsection{Data Structures and Optimization}
We represent our input graphs and coarse graphs in-memory using the compressed-sparse-row (abbreviated as CSR or CRS) format. We require a data structure to track connectivity of each vertex to each partition in order to facilitate Jet's iterations. Our label propagation iterations must be able to quickly identify the first and second most connected parts for each vertex. Our weak rebalancing iteration must identify the most connected valid destination part for each vertex in an oversized part. Our strong rebalancing iteration must sum the connectivity among valid destination parts for each vertex in an oversized partition. Finally, it should be possible to modify this data structure given a list of vertices to move. A naive implementation might use $|V|*k$ space to explicitly track this connectivity data for each possible pair of a vertex and part. Unfortunately, this uses far too much space with otherwise reasonable values for $k$, and is inefficient to traverse in all use cases. Our implementation is based on the observation that for any vertex $v$, the number of partitions to which it can have nonzero connectivity is at most $\textnormal{min}(k, \textnormal{degree}(v))$. We utilize a formulation similar to the CSR graph format to represent the vertex-part connectivity matrix. Our data structure allocates space equal to the following equation:
\begin{equation}
    |V| + 2\sum_{v \in V} \textnormal{min}(k, \textnormal{degree}(v))
\end{equation}
Each row in this CSR representation is treated as a hashtable (keyed on the partition id) for creation and updates. To determine the most connected parts that satisfy some filter criteria relative to each use case, we linearly search the hashtables. This linear search is substantially more efficient for smaller hashtables, so we limit the number of empty entries. Although $min(k, degree(v))$ is the maximum possible part connections for each vertex, we observe that many graphs (particularly regular graphs but even many irregular graphs) have a much smaller number of non-zero connections in practice. For instance, it is possible for a degree 100 vertex with $k = 128$ to only have one or two nonzero part connections. We set the hashtable size to be slightly larger than the initial connectivity upon construction. This may cause insertions into the hashtable to fail once this limited capacity is reached. When this occurs, we expand the hashtable capacity and recalculate its contents. We assign a small amount of extra space to each hashtable to limit the frequency with which this is necessary.
\\
\begin{algorithm}[htbp]
\caption{Jet - Update Part Connectivities.}
\label{alg:jetUpdate}
\begin{algorithmic}[1]
\Require The graph $G = (V, E)$. A partition array $P_s$. Data structures $DS$ for querying vertex-part connection info and lock status. A list of vertex moves $M$.
\Ensure Updated Data structures $DS$
\For{$v \in$ vertexSet($M$) in parallel}
    \State $p_s \gets P_s[v]$
    \For{$(u,w) \in E[v]$ in parallel}
        \State $h \gets $ getHashmap($DS, u$)
        \State $h[p_s]$ $\gets$ $h[p_s]$ - $w$
        \If{$h[p_s] = 0$}
            \State setOpen($h, p_s$)
        \EndIf
    \EndFor
\EndFor
\For{$(v, p_d) \in M$ in parallel}
    \For{$(u,w) \in E[v]$ in parallel}
        \State $h \gets $ getHashmap($DS, u$)
        \If{$p_d \notin h$}
            \State insert($h, p_d$)
        \EndIf
        \State $h[p_d]$ $\gets$ $h[p_d]$ + $w$
    \EndFor
\EndFor
\end{algorithmic}
\end{algorithm}
In order to update this data structure once the vertex move list $M$ is chosen, we update in two passes (see Algorithm \ref{alg:jetUpdate}). The first pass decrements the part connectivity of every neighbor of each vertex in $M$ for the respective source partitions, and creates an open entry in place of any part that reaches a connectivity of zero. The second pass increments the part connectivity of every neighbor of each vertex in $M$ for the respective destination partitions, potentially creating new entries in the hashtable when necessary. The creation of new entries in the second pass may fail for some rows if the current hashtable size for that row is insufficient. We mark the respective rows and then recalculate the corresponding hashtables in a third pass (not shown in Algorithm \ref{alg:jetUpdate}). All passes leverage atomic operations to ensure correctness, but a race condition affects which entry in the hashtable any given part will be assigned to. This race condition also exists for the initial construction of the data structure. In the Jetlp and Jetrw phases, this can affect how ties are broken when determining the most connected part for a vertex. Together with the race condition for bucket insertions in Jetrw and Jetrs, these are the only sources of non-determinism in the Jet refinement algorithm. Algorithm \ref{alg:jetUpdate} has the benefit of only updating rows adjacent to the vertex moves, and only the entries specifically affected in those rows. We also implement an alternative update algorithm, which we use when the number of vertex moves constitutes more than 10\% of the total vertices in the graph. In this alternative algorithm, we reconstruct the entire hashtable for every row adjacent to a moved vertex from the new partition state after applying each move. This reduces the irregularity of memory accesses over Algorithm \ref{alg:jetUpdate}, at the cost of more work.

\section{Experimental Setup}
Our experiments evaluate the performance of our partitioner in terms of both cutsize and overall execution time. We compare our GPU partitioner to other state-of-the-art multicore multilevel partitioners including Mt-Metis v0.7.2 with Hill-Scanning, mt-KaHIP v1.00 with MLS, KaMinPar v1.0, and Mt-KaHyPar-D v1.3.2, as well as the serial partitioner Metis v5.1.0. We choose to compare to Mt-Metis with Hill-Scanning enabled and to mt-KaHIP with MLS enabled because these are the highest quality refinement options available for their respective partitioners. We utilize the default configuration of Mt-KaHyPar as this is the highest quality configuration to use KFM refinement in v1.3.2. We are unable to compare with either other GPU partitioner \cite{FaggingerAuer2013, GKSG-GPU}, as their code is not available. In the later work \cite{GKSG-GPU}, their cutsize results were slightly worse than both Metis and Mt-Metis (without Hill-Scanning) on all graphs tested. We evaluate on $k=32$, $k=64$, $k=128$ and $k=256$ with the imbalance set to $3\%$, as well as $k=128$ with imbalance set to $1\%$ and $10\%$. This constitutes a total of six experiments per graph and partitioner. Although most of these partitioners can operate on arbitrary values of $k$, mt-KaHIP cannot; therefore, our experiments are on $k$ values that are powers of 2. For each combination of graph, experiment, and partitioner, we collect the median cutsize and median runtime across a number of runs. The number of runs performed is dependent on the partitioner: we perform five runs for mt-KaHIP MLS, 11 runs for KaMinPar, Mt-Metis HS, and Mt-KaHyPar-D, three runs for Metis, and 21 runs for our partitioner. The trials per partitioner are approximately inversely proportional to their respective runtimes. We present breakdowns versus each opposing partitioner by experiment configuration and in terms of graph classification.

\subsection{Effectiveness Tests}
In order to determine the effectiveness of our refinement method, we compare directly with mt-KaHIP's MLS and Mt-KaHyPar-D's KFM. As these are the highest quality partitioners to which we compare, these serve as ideal benchmarks for Jet refinement. In order to isolate refinement as the only variable, we export the coarse graph hierarchy and initial partitioning from the opposing partitioner and import it into our program. We then refine the solution on the imported hierarchy using Jet refinement. For each run, we compute the ratio of the final cutsize result obtained by the two refinement methods. We gather the median ratio out of a number of test runs for each graph (5 vs. MLS, 11 vs. KFM). We do the same for the refinement time. We also perform the reverse of this experiment, exporting our coarse graphs and initial partitioning into the opposing partitioner. To ensure a fair comparison, we exclude coarse graphs which have an imbalanced partition prior to refinement and export the partitioning for the coarsest graph with a balanced partition. This is necessary because MLS assumes a balanced input partition. For these experiments, we use $k=64$ and the imbalance equals $3\%$, and use our CPU platform to produce a fair refinement time comparison.

\subsubsection{Runtime}

We present a breakdown of our runtimes into three categories: coarsening, initial partitioning, and refinement. We analyze the runtime scaling versus $k$ and the imbalance. We analyze the runtime scaling versus graph size using several graph families. We present multicore scaling numbers and GPU vs CPU performance.

\subsection{Test Graphs}

Our test set (see Table~1 of the supplementary file) contains all graphs with at least 50 million nonzeroes but less than 750 million nonzeroes from the Suitesparse graph repository \cite{SuiteSparse} (excluding mawi graphs). We also include a few miscellaneous graphs (ppa, citation, products) from Open Graph Benchmark \cite{OGB} and some social networks (dblp10, amazon08, hollywood11, enwiki21) published by the Laboratory for Web Algorithmics \cite{BoVWFI, BRSLLP} and one graph (fe\_rotor) from the Walshaw Graph Benchmark \cite{walshaw}. We also add a 2000x4000 rectangular mesh (grid) and a 200x200x200 cubic mesh (cube). We preprocess all graphs by performing the following steps: we remove self-loops, convert all directed edges to undirected edges, remove duplicate edges, and extract the largest connected component. The graphs are further grouped into one of nine classes.

\subsection{Test Systems}
We conduct our tests on two different systems. Our first system runs on a 32-core Ryzen 3970x Threadripper, with 256GB of RAM (quad-channel DDR4). The first system runs the experiments for the competing partitioners, as well as the serial and multicore experiments for our partitioner's scaling results. Our MLS vs. Jet refinement effectiveness test also runs on this system. We run each partitioner using 64 threads. The second system is a virtual machine with 12 virtual cores of an Intel Xeon Gold 6342 CPU, 90GB RAM, and an Nvidia A100 GPU with 80GB of VRAM. The second system runs the primary experiments for our partitioner, and the Jetlp component effectiveness experiments. Both systems run Ubuntu 20.04. Our code is compiled with NVCC using Cuda Toolkit version 11.6.2 for the A100 platform, and g++ version 10.2.0 on the ThreadRipper 3970x platform. We use release versions 3.6.1 of both Kokkos and Kokkos-Kernels libraries, and Metis library version 5.1.0.

\section{Partitioner Performance Evaluation}

\subsection{Quality}
\begin{figure*}
    \centering
    \includegraphics[width=.49\textwidth]{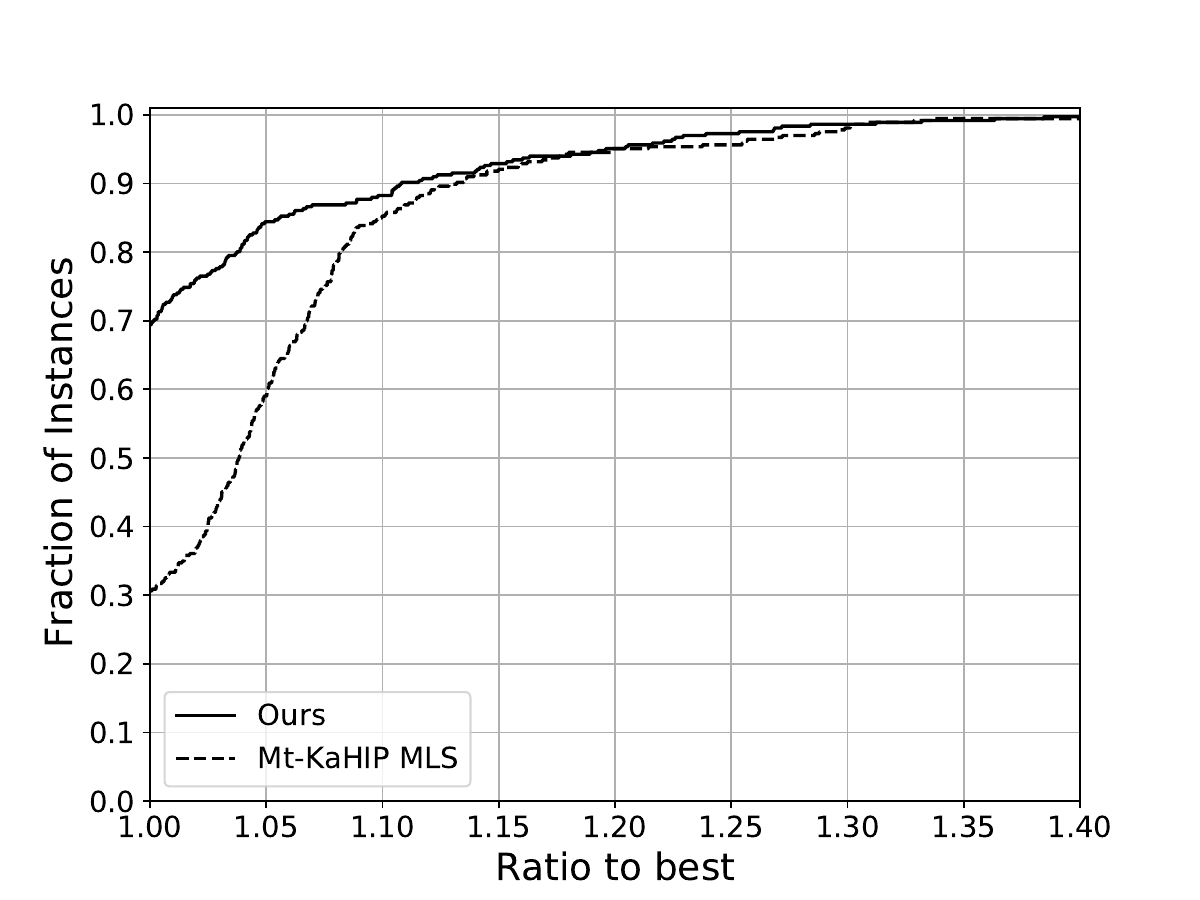}
    \includegraphics[width=.49\textwidth]{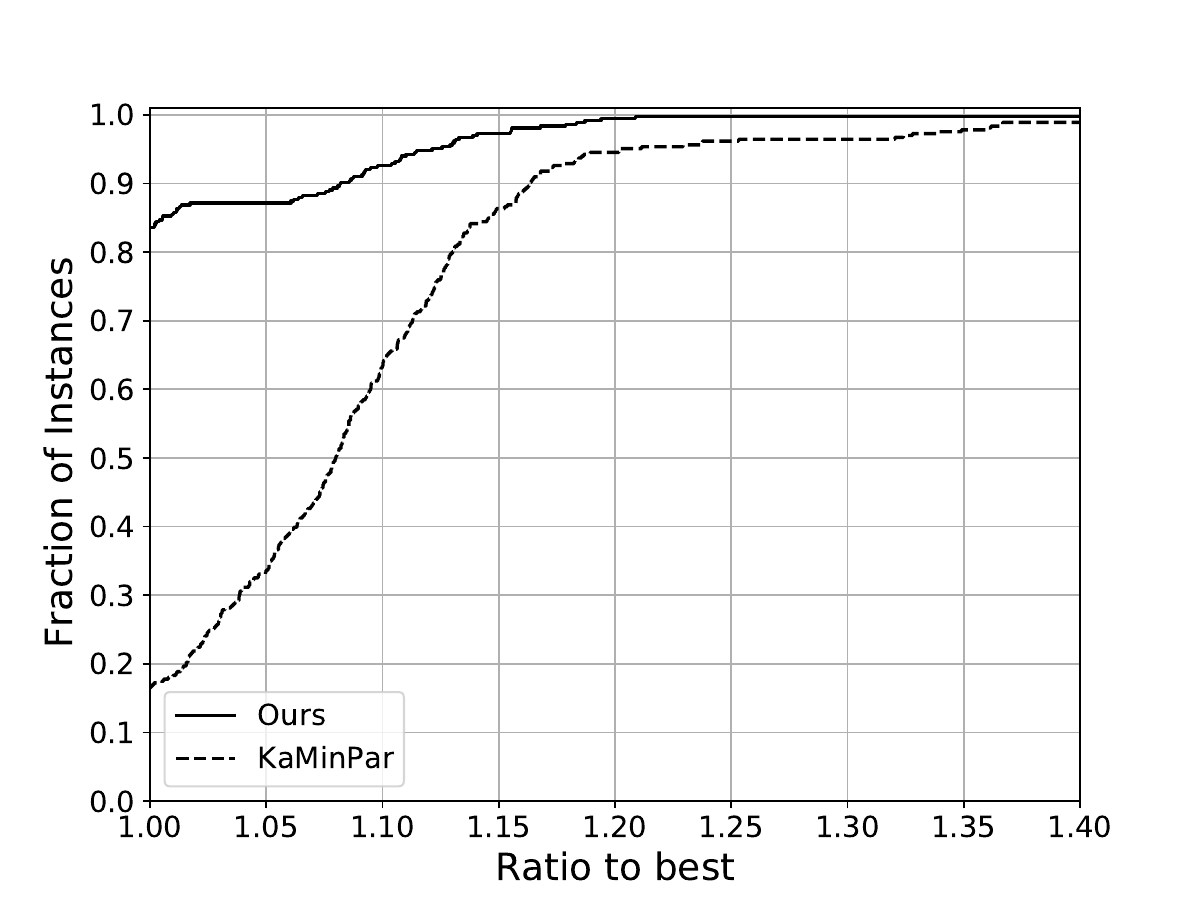}
    \includegraphics[width=.49\textwidth]{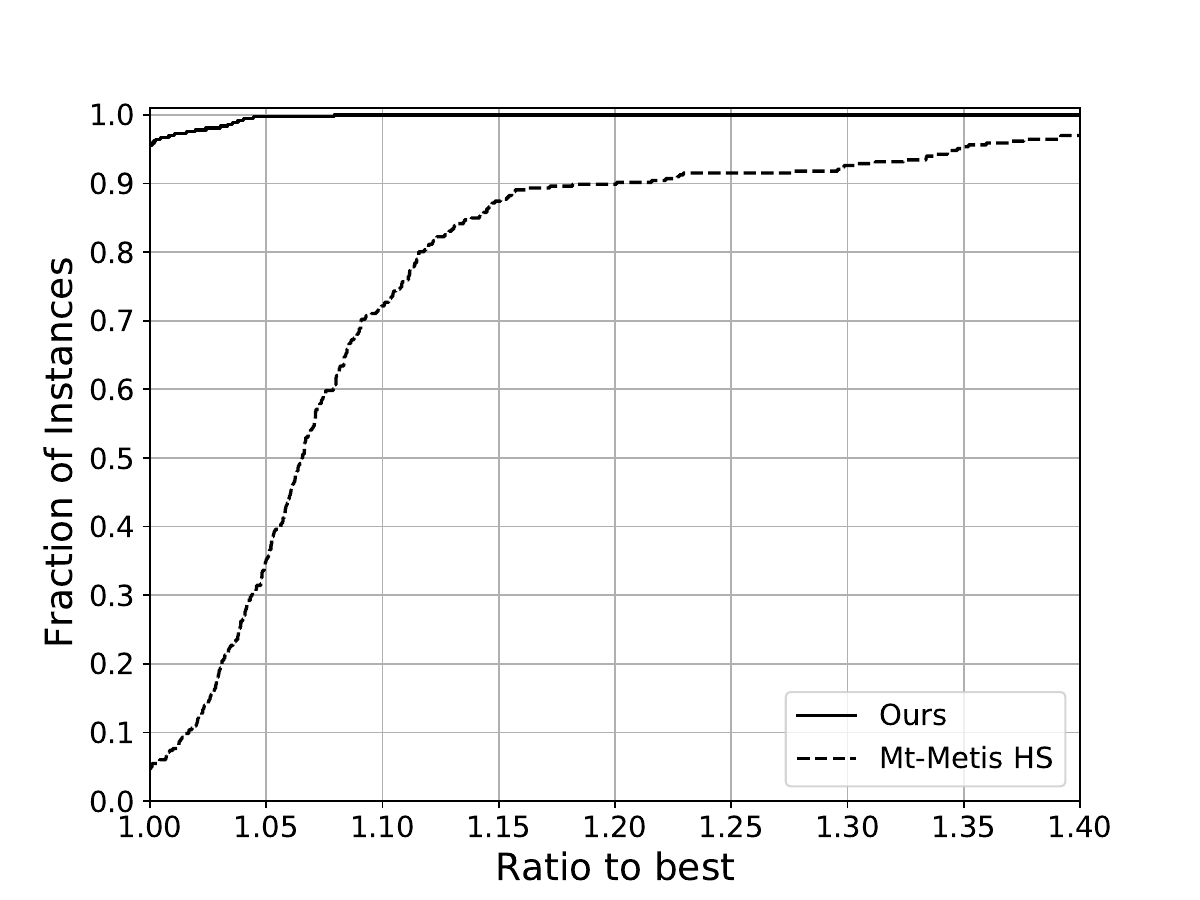}
    \includegraphics[width=.49\textwidth]{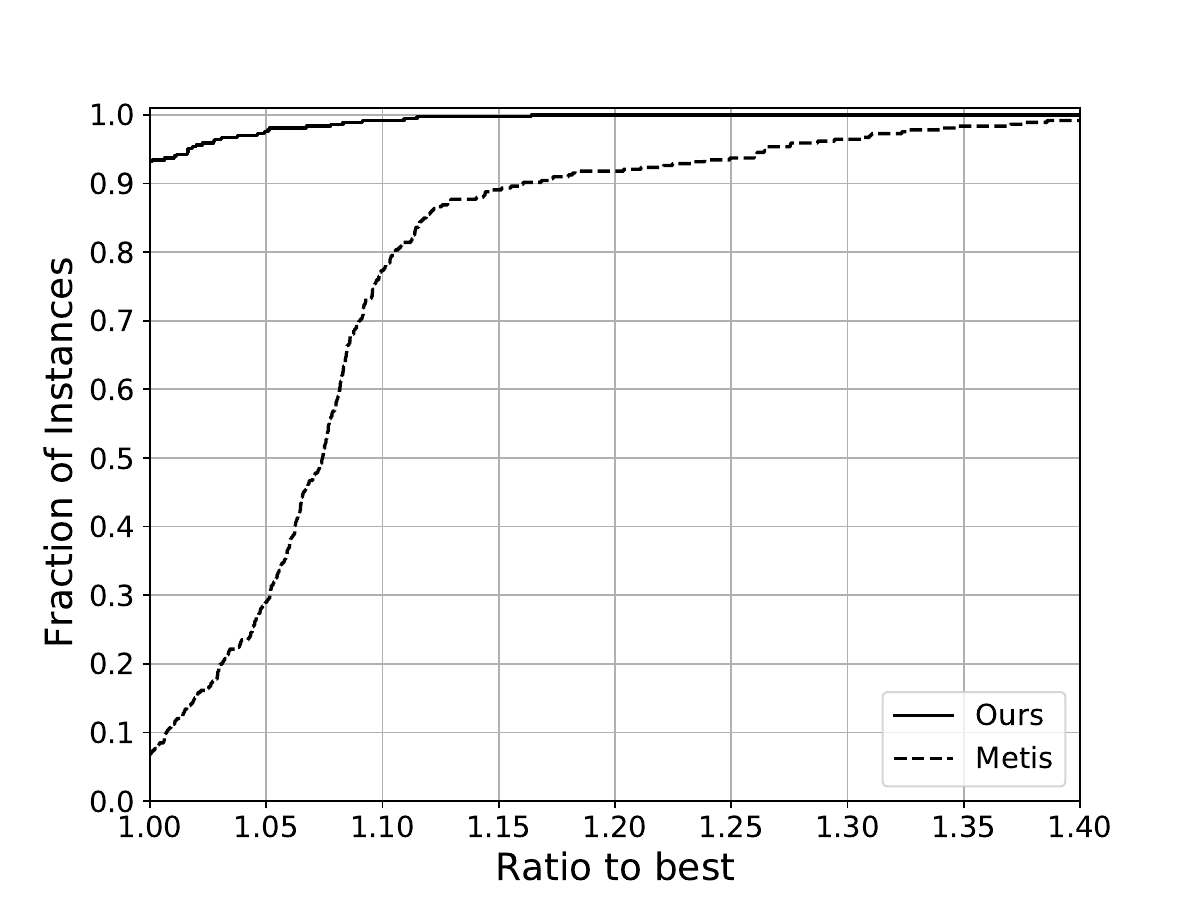}
    \includegraphics[width=.49\textwidth]{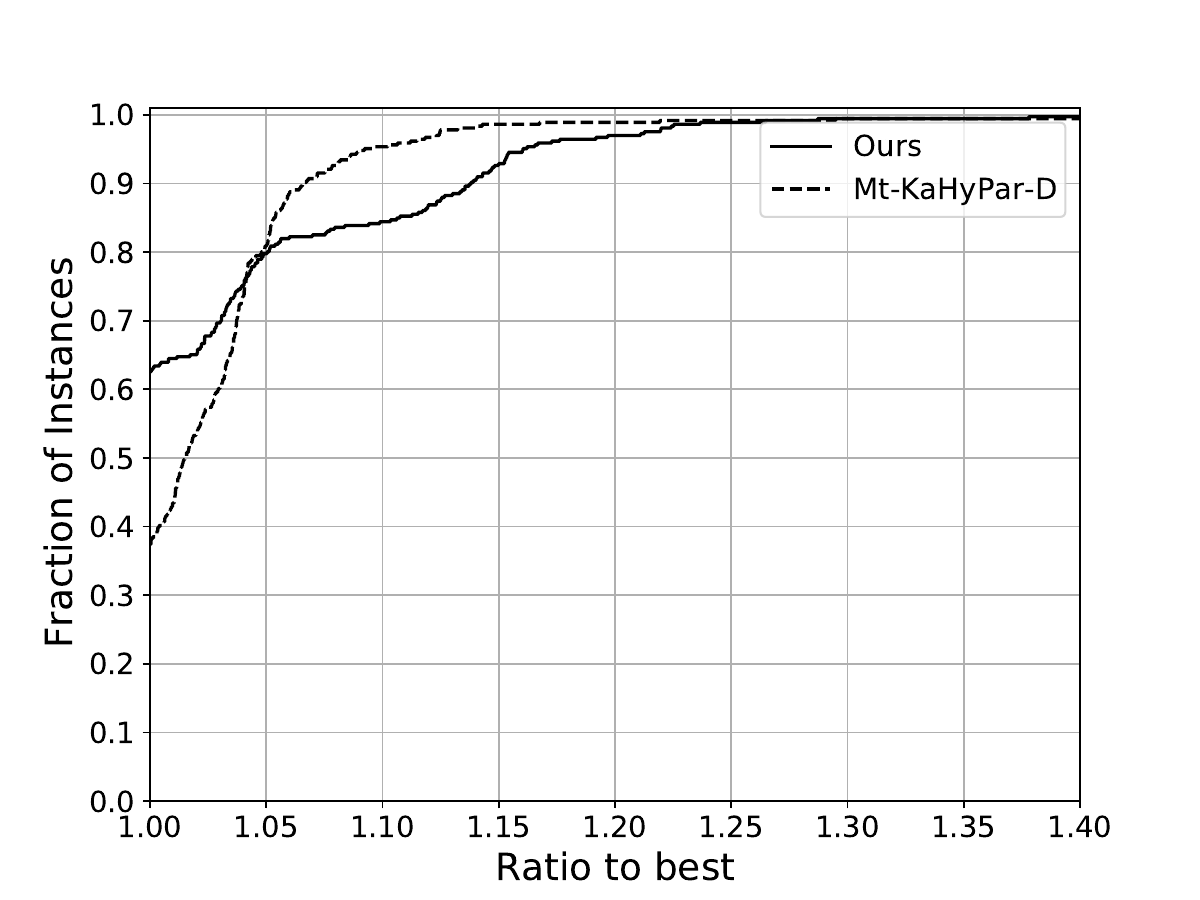}
    \caption{We use performance profiles to compare cutsize obtained using our partitioner to others.}
    \label{fig:cut_profiles}
\end{figure*}

Our GPU partitioner outperforms Mt-Metis with Hill-Scanning (HS), KaMinPar, and Metis in cutsize. As shown in Figure \ref{fig:cut_profiles}, our partitioner is better on more than 90\% of test instances than Mt-Metis HS and Metis, more than 80\% of instances versus KaMinPar, about 70\% of instances versus mt-KaHIP MLS, and more than 60\% of instances versus Mt-KaHyPar-D.

\subsubsection{Experiment Configs}
In Table \ref{tab:comp_by_k}, we note that our cuts are more than 6\% better than Mt-Metis HS, KaMinPar, and Metis in cutsize across all experiment configurations except one. The outlier is the $k=256$ experiment, where ours is 4.9\% better than KaMinPar. Mt-KaHIP MLS produces cuts around 2.5\% worse than ours overall. The only competitor to outperform ours is Mt-KaHyPar-D, which achieves 0.5\% better cuts overall. We note that our cutsize performance at $k=128$ and imbalance of $10\%$ is relatively better versus each other partitioner than the $k=128$ and imbalance of $3\%$ configuration. At $k=128$ and an imbalance of $1\%$, we are relatively better versus mt-KaHIP MLS and KaMinPar than in the respective $3\%$ configuration, whereas we are relatively worse versus Mt-Metis HS and Metis. We do relatively worse versus KaMinPar, Mt-KaHyPar-D, Mt-Metis HS, and Metis with increasing values of $k$.

\subsubsection{Graph Classes} \label{sec:part_class_eval} 
\begin{figure*}
\centering
\begin{subfigure}{0.49\textwidth}
    \includegraphics[width=\textwidth]{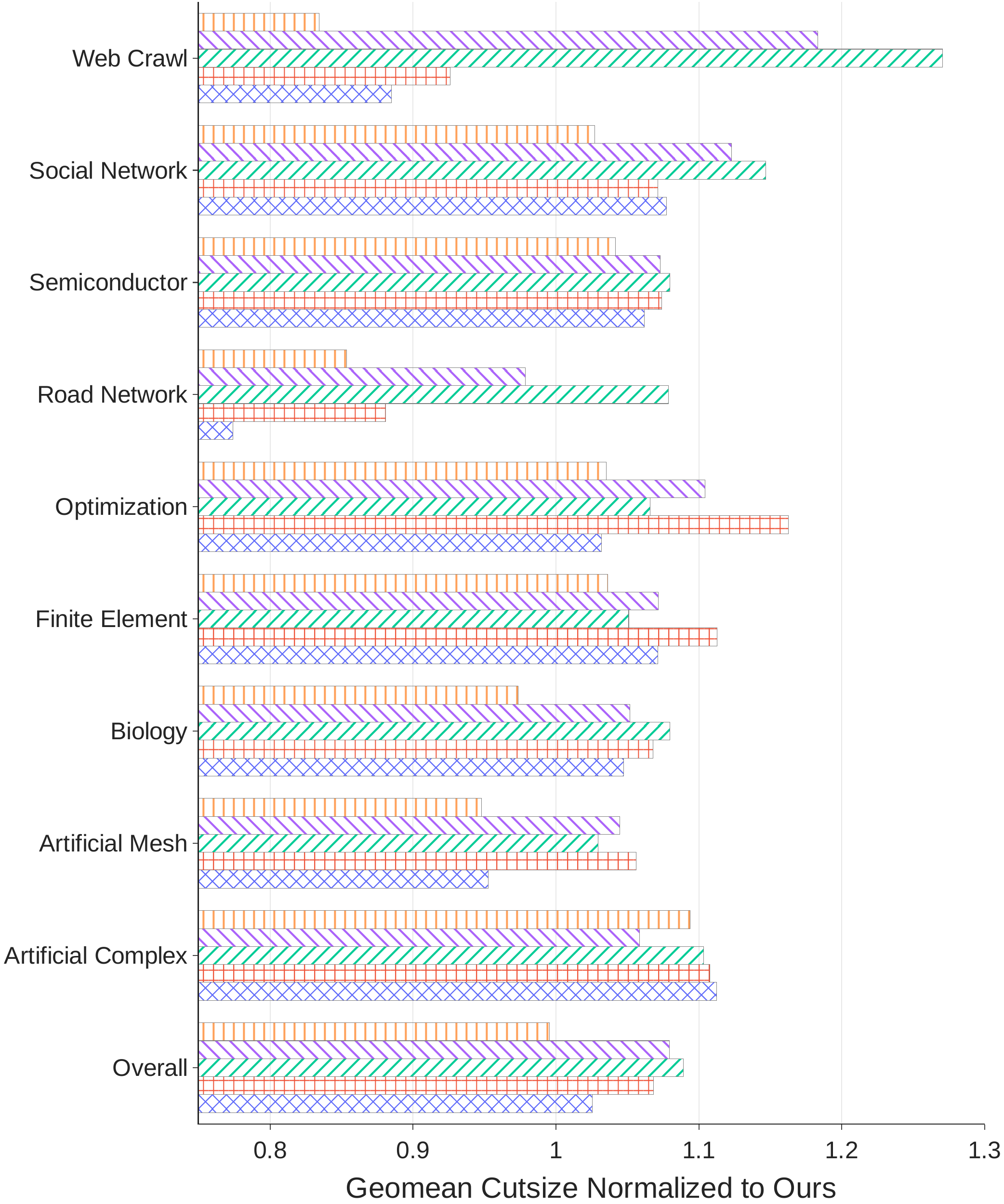}
    \caption{}
    \label{fig:cut_class}
\end{subfigure}
\begin{subfigure}{0.49\textwidth}
    \includegraphics[width=\textwidth]{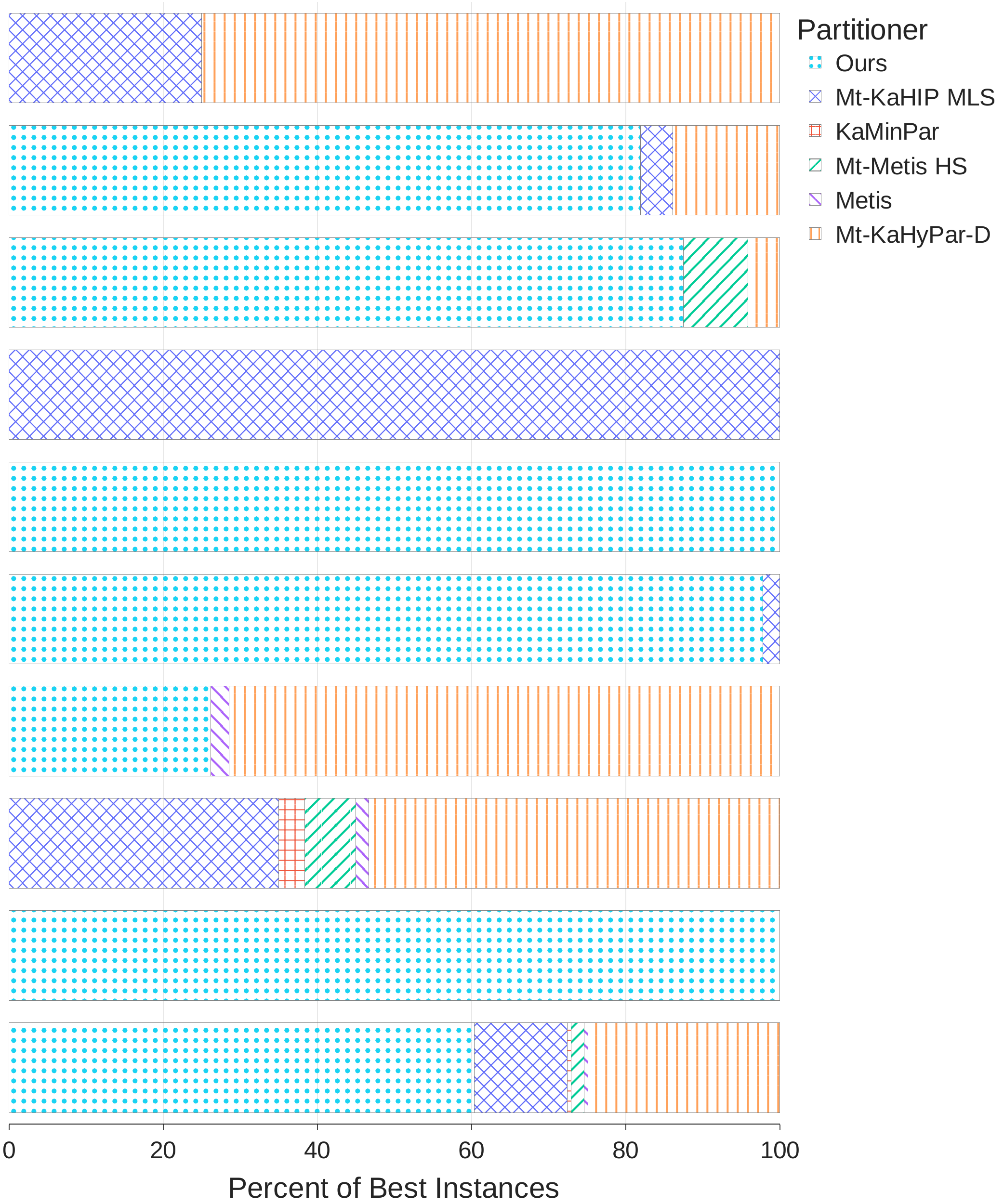}
    \caption{}
    \label{fig:best_in_class}
\end{subfigure}
\caption{Cutsize Results By Class}
\end{figure*}

We classify our partitioner's strengths and weaknesses by graph type using Figures \ref{fig:cut_class} and \ref{fig:best_in_class}. Our partitioner is dominant on finite element problems, optimization problems, social networks, semiconductor problems, and artificial complex networks. Of the social networks, our partitioner only failed to produce the best cut on all amazon08 and dblp10 instances, which are the two smallest social networks in our test set, and one instance for com-Orkut. We have a moderate strength on biology graphs, with ours obtaining the best cuts for most ppa and cage15 instances, and Mt-KaHyPar-D obtaining the best cuts on most of the kmer graph instances. Our weaknesses include the artificial meshes, web crawls, and road networks. Excluding the web crawls, most of the graphs in these classes have an underlying 2D structure. We explore the reason for our poor performance on the web crawls and 2D problems in the effectiveness test section. 

\begin{table*}[]
{   \scriptsize
    \centering
    \caption{We compare the Jet partitioner to various partitioners, reporting the ratios of the geometric mean of median cutsizes obtained with the partitioner to the geometric mean of the median cutsizes with the Jet partitioner. A value greater than 1 indicates that the Jet partitioner performs better. The number of parts and the balance constraint setting are varied.}
    
    \begin{tabular}{@{}lcccccc@{}}
        \toprule
        & $k = 32$ & $k = 64$ & $k = 128$ & $k = 256$ & $k = 128$ & $k = 128$  \\
        \emph{Partitioner} & i=3\% & i=3\% & i=3\% & i=3\% & i=1\% & i=10\% \\
        \cmidrule(l){2-7}
        mt-KaHIP MLS & 1.020 & 1.020 & 1.022 & 1.021 & 1.043 & 1.026 \\
        KaMinPar & 1.084 & 1.074 & 1.063 & 1.049 & 1.067 & 1.073 \\
        Mt-Metis HS & 1.111 & 1.094 & 1.084 & 1.073 & 1.075 & 1.100 \\
        Metis & 1.099 & 1.085 & 1.072 & 1.063 & 1.069 & 1.088 \\
        Mt-KaHyPar-D & 0.995 & 0.997 & 0.994 & 0.991 & 0.995 & 1.000 \\
        \bottomrule
    \end{tabular}
    \label{tab:comp_by_k}
}
\end{table*}

\subsection{Runtime}
\begin{figure*}
    \centering
    \includegraphics[width=.49\textwidth]{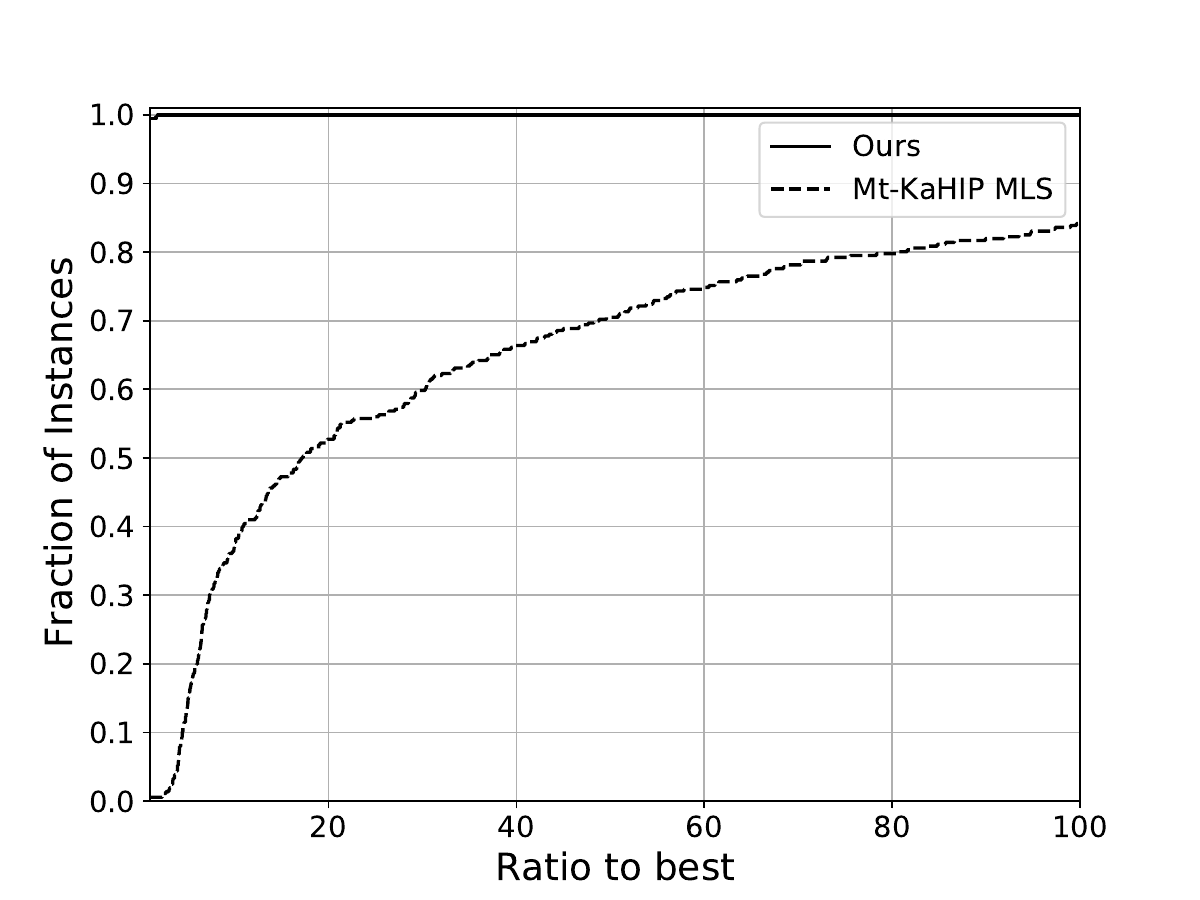}
    \includegraphics[width=.49\textwidth]{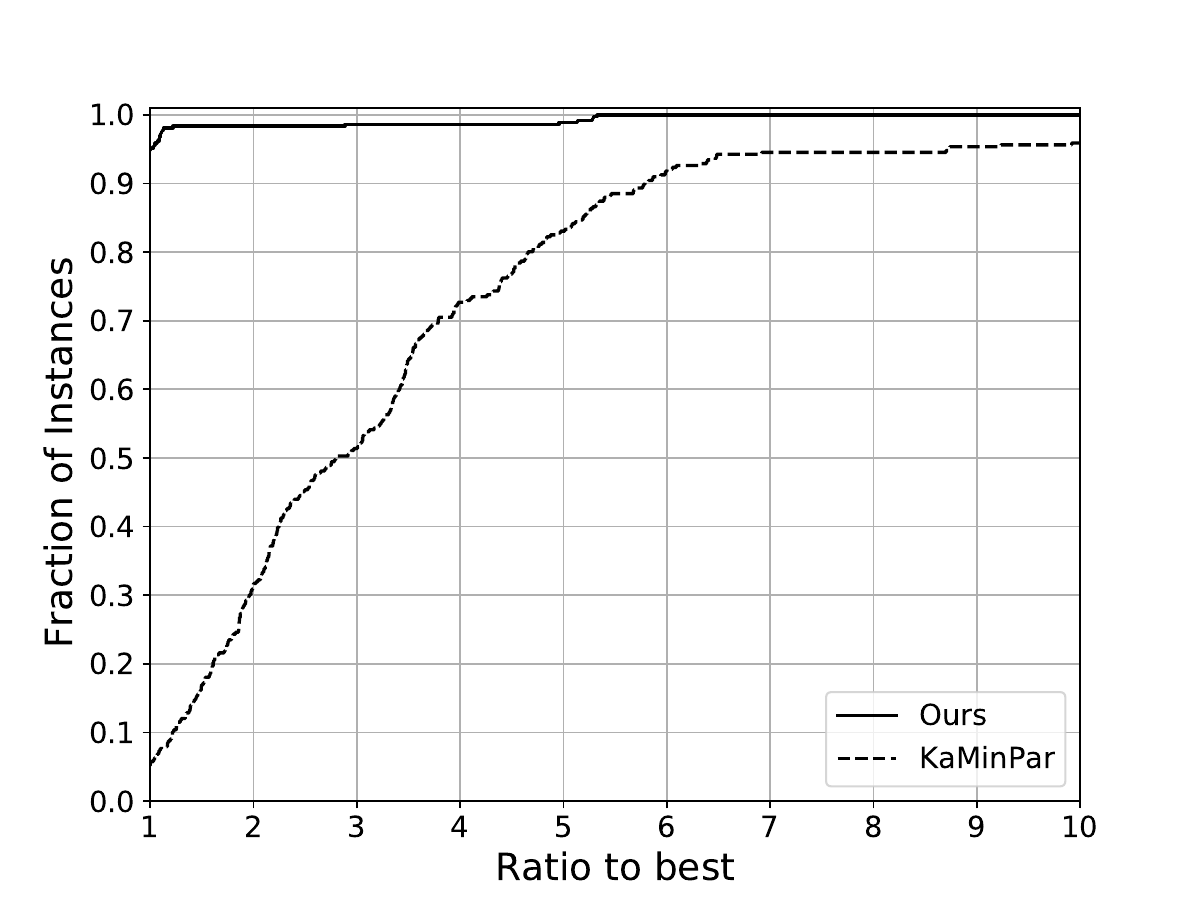}
    \includegraphics[width=.49\textwidth]{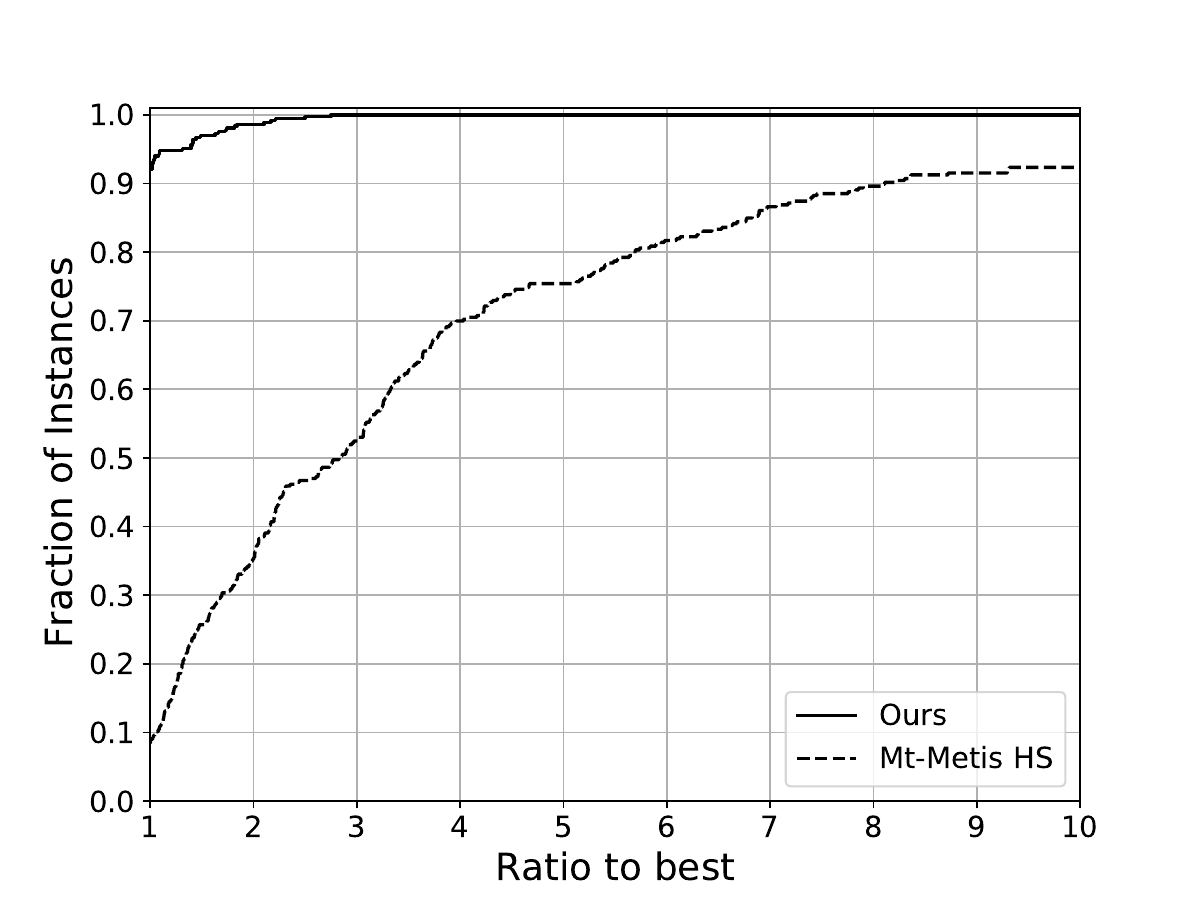}
    \includegraphics[width=.49\textwidth]{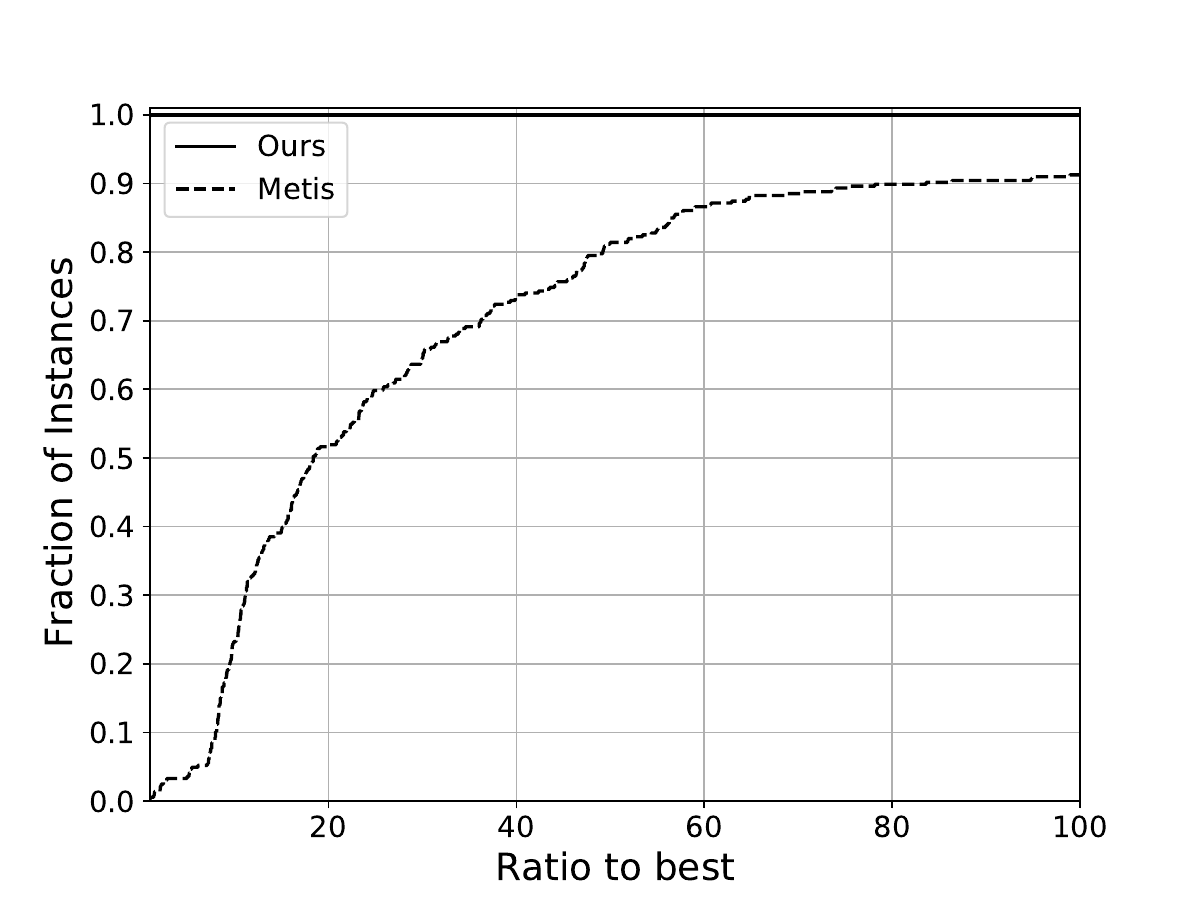}
    \includegraphics[width=.49\textwidth]{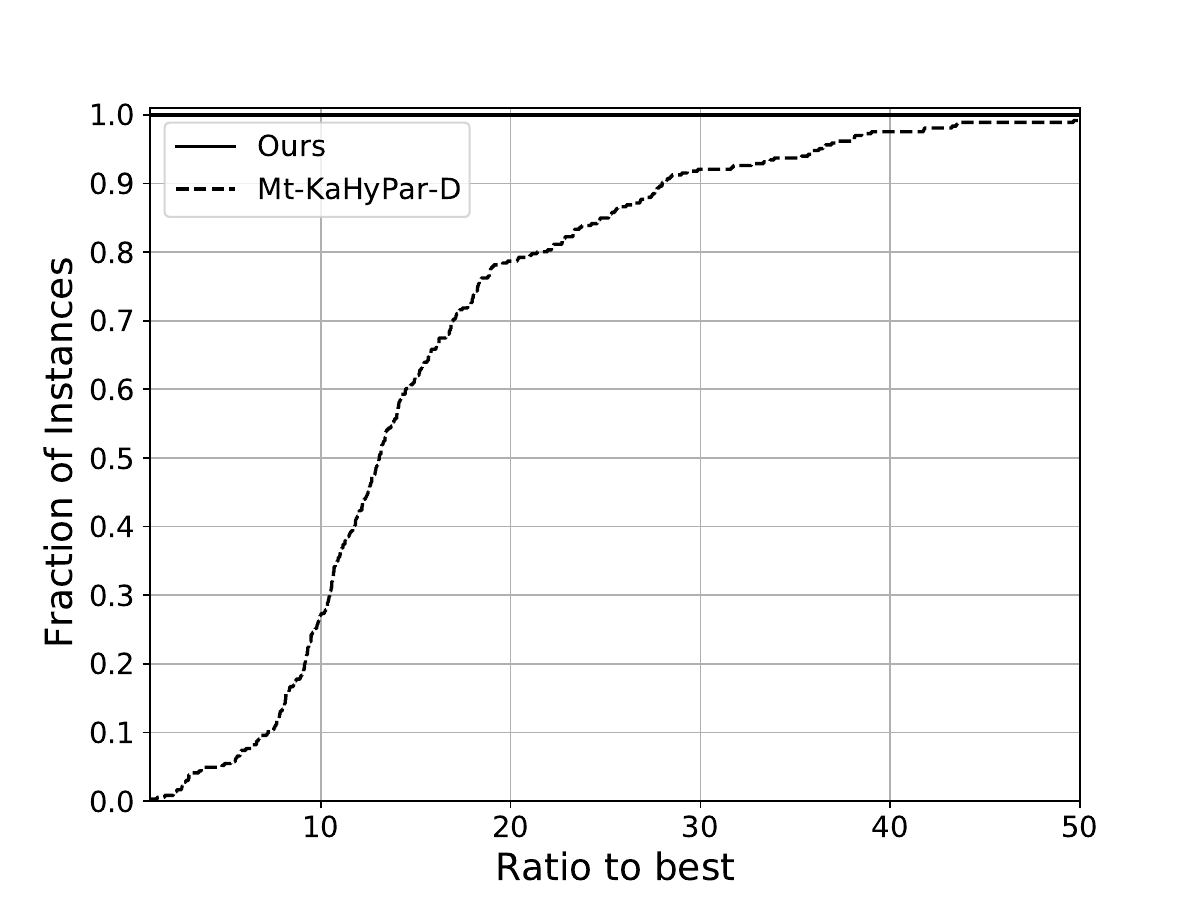}
    \caption{We use performance profiles to compare partitioning time of the Jet partitioner (Ours) on the A100 GPU to the execution time of other partitioners. The other partitioners are executed on the AMD Ryzen Threadripper 3970x CPU.}
    \label{fig:time_profiles}
\end{figure*}

\begin{figure*}
\centering
\begin{subfigure}{0.49\textwidth}
    \includegraphics[width=\textwidth]{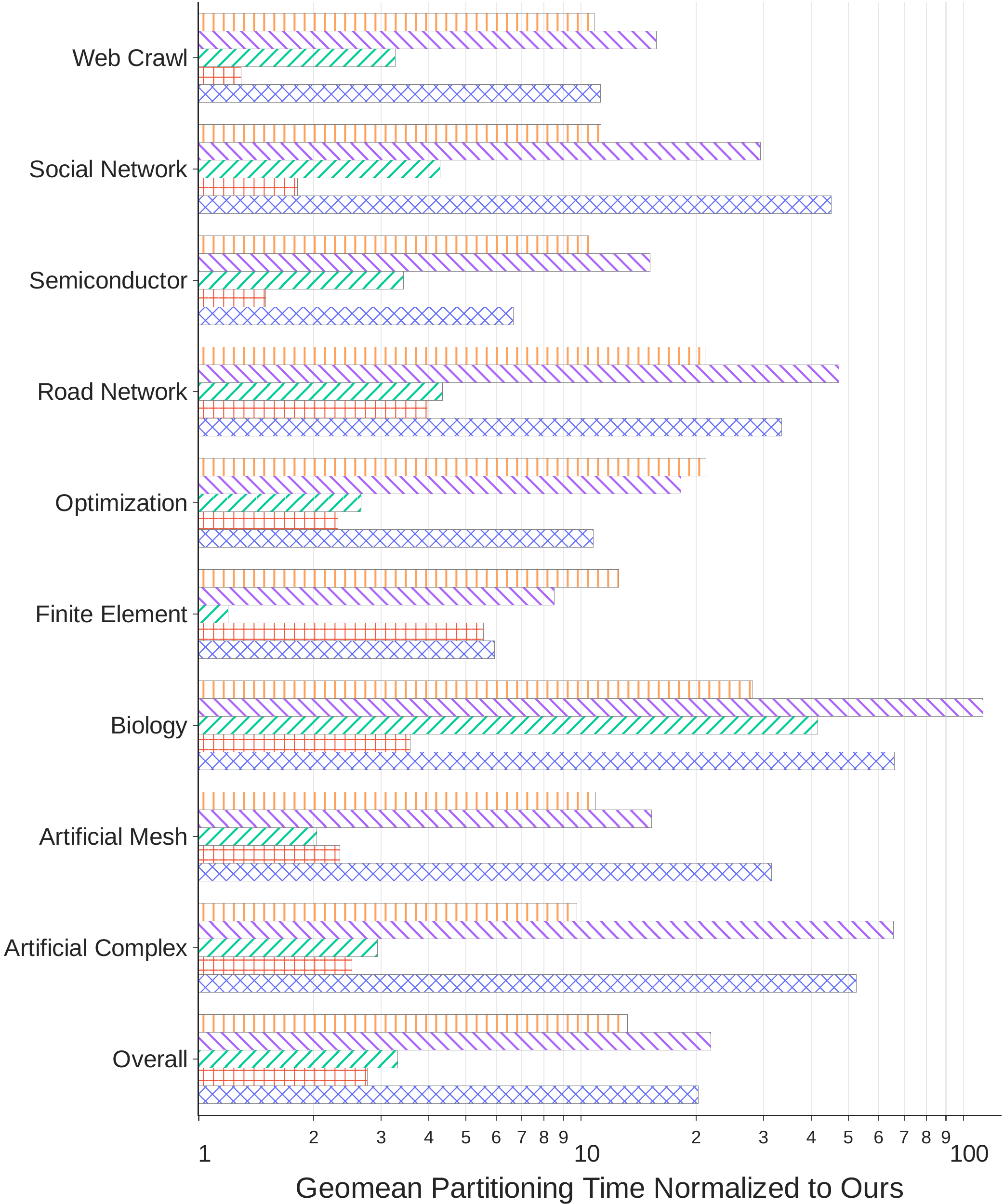}
    \caption{By Graph Class}
    \label{fig:parttime_class}
\end{subfigure}
\hfill
\begin{subfigure}{0.49\textwidth}
    \includegraphics[width=\textwidth]{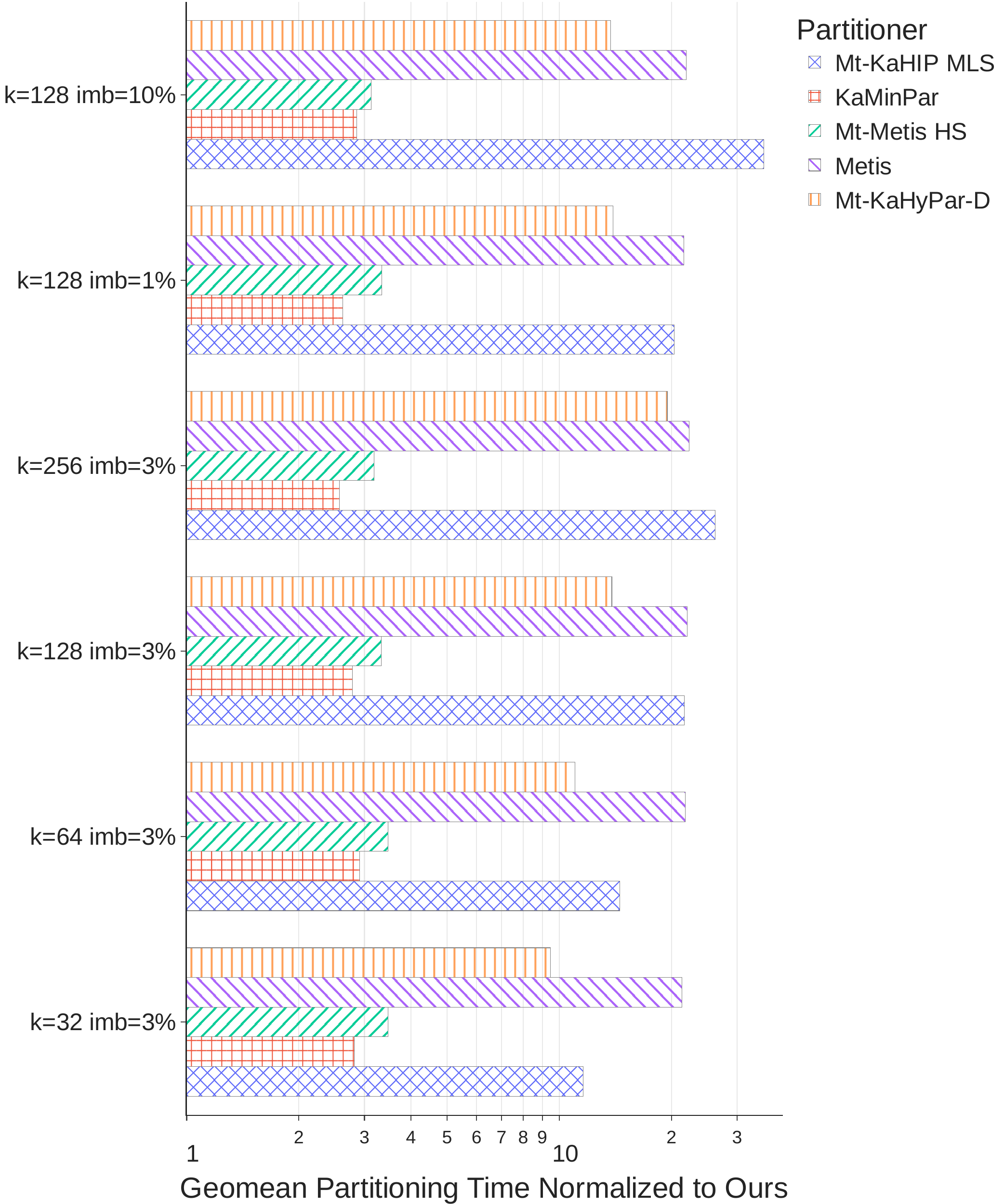}
    \caption{By Experiment}
    \label{fig:parttime_exp}
\end{subfigure}
\caption{Partitioning Time Comparison}
\end{figure*}

Our GPU partitioner is consistently faster than our CPU competitors, with shorter runtimes than any competitor in more than 85\% test instances. We found that our partitioner was faster than mt-KaHIP MLS on more than 99\% test instances (see Figure \ref{fig:time_profiles}), and more than twenty times faster on more than 40\% test instances. Compared to KaMinPar, Figure \ref{fig:time_profiles} shows that ours is faster in more than 90\% instances and at least twice as fast in more than 65\% instances. Compared to Mt-Metis HS, our runtime was better in more than 90\% test instances and at least twice as fast in more than 60\% instances. Our partitioner is faster than Metis by more than 20x in over 40\% of test instances, similar to mt-KaHIP MLS. Our partitioner is faster than Mt-KaHyPar-D by at least 10x in over 70\% of test instances. The runtime performance of mt-KaHIP is surprising; in Figure \ref{fig:parttime_exp}, it only achieves superior geometric mean runtimes to Metis on our $k=32$ and $k=64$ experiments. In the same figure, our partitioner shows similar performance trends across experiment configurations. In Figure \ref{fig:parttime_class}, our partitioner achieves strong runtime results on artificial complex, biology, road network, semiconductor, and social network graphs. Our partitioner achieves similar runtime performance to Mt-Metis HS on the finite element graphs, and it also achieves similar performance to KaMinPar on the web crawl graphs. KaMinPar achieves much better performance than our partitioner on the circuit5M graph.

\subsubsection{Time Breakdown}
In Table \ref{tab:time_breakdown}, we present the average time spent in each subtask of partitioning by our graph partitioner on the GPU. We organize this data by graph class, and include the average total runtime among all experiments in that graph class. Across all classes, initial partitioning is responsible for at most 10.6\% of the total runtime. Coarsening dominates on artificial complex, semiconductor, and web crawl graphs, whereas uncoarsening dominates on the other graphs. Coarsening tends to dominate on most graphs where the degree distributions are extremely irregular, whereas uncoarsening dominates on the more regular graphs.

\begin{table}[]
{    \scriptsize
    \centering
    \caption{Partitioning time breakdown by subtask and average total time (seconds).}
    \begin{tabular}{@{}lS[table-format=2.1]S[table-format=2.1]S[table-format=2.1]S[table-format=1.2]@{}}
    \toprule
        \emph{Graph Class} & {Uncoarsen (\%)} & {Coarsen (\%)} & {InitPart (\%)} & {Avg. time (s)} \\
        \cmidrule(lr){2-4} \cmidrule(l){5-5}
        Web Crawl & 30.3 & 64.6 & 5.1 & 1.38 \\
Social Network & 58.2 & 31.2 & 10.6 & 1.11 \\
Semiconductor & 38.9 & 56.2 & 4.9 & 1.63 \\
Road Network & 52.6 & 43.8 & 3.6 & 0.43 \\
Optimization & 50.4 & 45.3 & 4.3 & 0.63 \\
Finite Element & 55.4 & 34.9 & 9.7 & 0.27 \\
Biology & 63.9 & 28.3 & 7.9 & 1.72 \\
Artificial Mesh & 64.6 & 31.2 & 4.2 & 0.41 \\
Artificial Complex & 36.9 & 54.7 & 8.4 & 1.83 \\
        \bottomrule
    \end{tabular}
    \label{tab:time_breakdown}
}
\end{table}

\section{Refinement Performance Evaluation}

\subsection{Effectiveness Test}
In our cut effectiveness test (table \ref{tab:effectiveness_cut}), we directly evaluate the performance of our Jet refinement algorithm. In Section~\ref{sec:part_class_eval} we found that our partitioner produces superior cuts on finite element problems, social networks, artificial complex networks, semiconductor problems, optimization problems, and some biology problems. Conversely, we found it to produce inferior cuts on problems derived from a 2D structure or from web crawls. Additionally, Mt-KaHyPar-D achieved superior cuts on the kmer graphs. We now detail how the refinement algorithm affects the partitioning results.

\subsubsection{Strengths} Overall, Jet produces 1.9\% better cuts than MLS when using the Mt-KaHIP coarsening and initial partitioning, and 6.2\% better cuts using our coarsening and initial partitioning. Compared to KFM, these numbers are 0.7\% with the Mt-KaHyPar-D backend and 3.4\% with our backend. Our refinement algorithm produces superior cuts for finite element, biology (excluding kmer graphs), social network, web crawl, and artificial complex networks. This rules out our refinement algorithm as the cause for our partitioner's weakness on web crawls, leaving our coarsening and initial partitioning as potential culprits. The cutsize result for semiconductor and optimization problems depends on the backend. Our backend produces the overall better cuts for most of these graphs, and Jet is superior in this case. Regarding kmer graphs, we found that Jet produces about 0.8\% better cuts than MLS and 1.4\% worse cuts than KFM combined across both backends. Strangely, Jet outperforms KFM on the road networks using our backend. However, our backend generates worse overall cuts for road networks than Mt-KaHIP or Mt-KaHyPar-D.

\subsubsection{Weaknesses} For problems with a 2D structure, including most artificial meshes and road networks, Jet demonstrates a refinement capability inferior to that of MLS and KFM. We speculate that this weakness is related to certain types of improvement that are difficult for our algorithm to identify. Improvements that substantially change the location of the boundary are difficult to find as our label propagation phase can only consider vertices that currently lie on the boundary within each iteration. This problem is exacerbated by graphs with large diameters, such as 2D-meshes. Road networks and artificial meshes (except the cubic mesh) have an underlying 2D structure and, therefore, have graph diameters of O($\sqrt{|V|}$). The kmer graphs also have large graph diameters. For a concrete example, consider our grid graph, which has a graph diameter of 5998: the MLS-to-Jet ratio is 0.902 and 0.906 for the mt-KaHIP backend and our backend, respectively. The cubic graph for comparison has a smaller graph diameter of 597 (a consequence of being a 3D mesh), and the MLS-to-Jet ratios are 0.965 and 1.008, respectively. MLS and KFM have the capability to find improvements that substantially shift the boundary, as they can perform long sequences of moves in localized regions of a graph. HS lacks this ability, due to the cap on hill-size.

\subsubsection{Uncoarsening Time} In Table \ref{tab:effectiveness_time}, Jet is faster than MLS for both backends by factors greater than 2.4x, and it is faster than KFM for both backends by at least 3x. Jet achieves these speedups consistently across most graph classes, except semiconductor graphs vs. MLS and artificial complex graph versus both competitors. We attribute Jet's superior uncoarsening speed to its bulk-synchronous design, efficient datastructures for tracking vertex-part connectivity, and the absence of priority queues. Although we have used the CPU platform for our code in order to obtain a fair comparison in this experiment, we note that a GPU implementation of either MLS or KFM is non-trivial.

\subsubsection{Component Effectiveness} In Table \ref{tab:lp_versions}, we evaluate the impact of design choices in our Jetlp phase compared to a baseline synchronous LP. Our baseline only moves vertices into their best connected partition, omits the afterburner kernel in its entirety, and ignores the lock bit. We compare four versions of the LP phase. The first alternate is the baseline plus vertex locking. The second alternate is the baseline plus a weaker version of the afterburner, which only considers vertex moves with positive or zero gain. The third alternate is the baseline plus the full afterburner, that is, it considers negative gain vertices as described in Section \ref{sec:neg_gain}. The fourth version is the full Jetlp algorithm, that is, the baseline plus vertex locking plus the full afterburner. The results in Table \ref{tab:lp_versions} show that the afterburner performs substantially better if it can consider negative gain vertex moves than when it does not. Interestingly, the vertex lock does not provide any benefit alone, but combined with the full afterburner it provides a benefit of 2.2\% versus the full afterburner without the locks. Full Jetlp provides a cutsize benefit over the baseline that varies from a negligible $0.1$\% for artificial complex graphs to a substantial $11.8$\% for artificial meshes. Furthermore, we investigate the impact of $\phi$ on the execution time and cutsize results of our final version. We found that decreasing our refinement tolerance value $\phi$ to 0.99 improves the uncoarsening time by 55\% and worsens the cutsize by 1.1\% over our default value of 0.999. Increasing $\phi$ to 0.9999 worsens the uncoarsening time by 34\% and improves the cutsize by 0.5\% over the default value. The cutsize benefit of increasing $\phi$ is most pronounced for artificial meshes and least pronounced for artificial meshes and web crawls.

\begin{table*}[]
{\scriptsize
    \centering
    \caption{Geomean(Baseline Cutsize) / Geomean(Version Cutsize).}
    \begin{tabular}{@{}cccc@{}}
    \toprule
        Baseline + Locks & Baseline + Weak Afterburner & Baseline + Full Afterburner & Full Jetlp \\
        \midrule
        1.000 & 1.009 & 1.030 & 1.052 \\
        \bottomrule
    \end{tabular}
    \label{tab:lp_versions}
}
\end{table*}

\begin{table}[]
{    \scriptsize
    \centering
    \setlength{\tabcolsep}{3pt}
    \caption{Refinement Effectiveness - Geomean of Median Ratio (Their Cut / Our Cut).}
    \begin{tabular}{@{}lcccc@{}}
    \toprule
         & MLS vs Jet & MLS vs Jet & K-way FM vs Jet & K-way FM vs Jet \\
        \emph{Graph Class} & Mt-KaHIP Backend & Our Backend & Mt-KaHyPar-D Backend & Our Backend \\
        \cmidrule(r){2-5}
        All & 1.019 & 1.062 & 1.007 & 1.034 \\
        Web Crawl & 1.019 & 1.100 & 1.004 & 1.043 \\
        Social Network & 1.082 & 1.173 & 1.040 & 1.079 \\
        Semiconductor & 0.998 & 1.063 & 1.006 & 1.029 \\
        Road Network & 0.905 & 0.901 & 0.972 & 1.024 \\
        Optimization & 0.985 & 1.013 & 0.998 & 1.010 \\
        Finite Element & 1.019 & 1.037 & 1.000 & 1.021 \\
        Biology & 1.034 & 1.059 & 0.995 & 1.023 \\
        Artificial Mesh & 0.934 & 0.977 & 0.951 & 0.986 \\
        Artificial Complex & 1.148 & 1.167 & 1.132 & 1.122 \\
        \bottomrule
    \end{tabular}
    \label{tab:effectiveness_cut}
    }
\end{table}

\begin{table}[]
{    \scriptsize
    \centering
    \setlength{\tabcolsep}{3pt}
    \caption{Refinement Effectiveness: Geomean of Median Ratio (Their Uncoarsening Time / Our Uncoarsening Time).}
    \begin{tabular}{@{}lcccc@{}}
    \toprule
         & MLS vs Jet & MLS vs Jet & K-way FM vs Jet & K-way FM vs Jet \\
        \emph{Graph Class} & Mt-KaHIP Backend & Our Backend & Mt-KaHyPar-D Backend & Our Backend \\
        \cmidrule(l){2-5}
        All & 2.418 & 2.498 & 3.005 & 3.328 \\
        Web Crawl & 1.713 & 2.457 & 3.455 & 4.806 \\
        Social Network & 2.827 & 2.188 & 2.395 & 2.197 \\
        Semiconductor & 0.997 & 1.291 & 2.684 & 3.372 \\
        Road Network & 3.083 & 3.271 & 4.787 & 5.961 \\
        Optimization & 2.165 & 2.114 & 3.868 & 4.806 \\
        Finite Element & 2.109 & 2.214 & 3.768 & 4.424 \\
        Biology & 5.335 & 6.379 & 3.562 & 3.892 \\
        Artificial Mesh & 1.759 & 1.519 & 2.903 & 3.005 \\
        Artificial Complex & 0.867 & 1.291 & 1.314 & 1.516 \\
        \bottomrule
    \end{tabular}
    \label{tab:effectiveness_time}}
\end{table}

\subsection{Parallel Scaling}
We continue the performance analysis by analyzing the relative performance of our test systems. In Table \ref{tab:ref_speedup}, we compare the performance of our multicore AMD Ryzen 3970x CPU system to our Nvidia A100 GPU system, and also compare multicore performance to serial performance on the CPU. We include results for total uncoarsening time, as well as refinement time for just the finest level graph. The 32-core uncoarsening speedup is between 7.8x and 16.1x, which is not ideal on the lower end. The finest level refinement speedup is within a similar range of 7.8x to 16x. The suboptimal 32-core speedup is partly due to memory-bandwidth constraints, as well as certain implementation choices made for GPU performance that aren't as effective for CPU platforms. The GPU vs. CPU uncoarsening speedup is between 3.5x and 9.2x. However, the GPU vs CPU finest level refinement speedup is better, landing between 6.3x and 14.8x. This is most likely due to the host-device synchronization time which represents a larger portion of the refinement time on the smaller coarse graphs.

\begin{table}[]
{    \scriptsize
    \centering
    \caption{Uncoarsening Speedup, Overall and Finest Level Refinement.}
    \begin{tabular}{@{}lS[table-format=1.2$\times$]
                        S[table-format=2.2$\times$]
                        S[table-format=2.2$\times$]
                        S[table-format=2.2$\times$]@{}}
    \toprule
         & \multicolumn{2}{c}{A100 vs 3970x 32-core} & \multicolumn{2}{c}{3970x 32-core vs Serial} \\
        \emph{Graph Class} & {Overall} & {Finest Level} & {Overall} & {Finest Level}\\
        \cmidrule(l){2-3} \cmidrule(l){4-5}
        Web Crawl & 5.14$\times$ & 7.99$\times$ & 9.28$\times$ & 9.72$\times$ \\
        Social Network & 7.71 & 9.94 & 12.45 & 12.82 \\
        Semiconductor & 6.04 & 8.18 & 7.83 & 7.77 \\
        Road Network & 6.53 & 14.77 & 7.95 & 8.46 \\
        Optimization & 5.48 & 9.04 & 13.04 & 10.44 \\
        Finite Element & 3.50 & 6.27 & 10.72 & 12.74 \\
        Biology & 9.22 & 13.14 & 14.04 & 11.71 \\
        Artificial Mesh & 4.33 & 8.96 & 9.72 & 10.61 \\
        Artificial Complex & 8.38 & 12.00 & 16.12 & 16.04 \\
        \bottomrule
    \end{tabular}
    \label{tab:ref_speedup}}
\end{table}

\section{Conclusion}
We demonstrate a partitioner that leverages GPU acceleration to decrease partitioning time whilst delivering state-of-the-art partition quality. Our partitioner demonstrates superior quality on five of nine graph classes in our test set compared to several state-of-the-art partitioners, across six experiment configurations. Our runtimes are superior on all nine graph classes. We attribute these results to our novel partition refinement algorithm, Jet. Jet builds on label propagation by addressing many common drawbacks while optimizing for GPU scalability. Jet delivers cutsizes similar to or better than two state-of-the-art parallel implementations of FM refinement on six out of nine graph classes, and superior runtime on seven out of nine graph classes. We identify quality on two-dimensional mesh-like graphs as the primary weakness of Jet, which is consistent with other label propagation algorithms. Our partitioner is able to substantially reduce the time spent for initial partitioning by coarsening to extremely small graphs. We plan to investigate methods to enhance Jet's quality and to demonstrate Jet in a distributed memory partitioner.

\section*{Acknowledgments}
Sandia National Laboratories is a multimission laboratory managed and operated by National Technology and Engineering Solutions of Sandia, LLC., a wholly owned subsidiary of Honeywell International, Inc., for the U.S. Department of Energy’s National Nuclear Security Administration under contract DE-NA-0003525.
This research was supported by the Exascale Computing Project (17-SC-20-SC), a collaborative effort of the U.S. Department of Energy Office of Science and the National Nuclear Security Administration.

\bibliographystyle{siamplain}
\bibliography{refs}

\end{document}